%% file: main_v3.2.tex
\documentclass[reprint, amsmath, amssymb, aip, superscriptaddress, floatfix]{revtex4-2}

\usepackage{graphicx}
\usepackage{dcolumn}
\usepackage{bm}
\usepackage{xcolor}
\usepackage[normalem]{ulem}
\usepackage{amsmath}
\usepackage[english]{babel}
\usepackage{braket}
\usepackage{units}
\usepackage[T1]{fontenc}
\usepackage{subfiles}

\newcommand{\q}{\hat{q}}

\usepackage{xr}
\makeatletter

\newcommand*{\addFileDependency}[1]{
\typeout{(#1)}
\@addtofilelist{#1}

\IfFileExists{#1}{}{\typeout{No file #1.}}
}\makeatother

\newcommand*{\myexternaldocument}[1]{%
\externaldocument{#1}%
\addFileDependency{#1.tex}%
\addFileDependency{#1.aux}%
}
\myexternaldocument{supplement}

\begin{document}

\preprint{APS/123-QED}

\title{Fast superconducting qubit control with subharmonic drives}

\author{Mingkang Xia}
\affiliation{Department of Physics and Astronomy, University of Pittsburgh, Pittsburgh, PA 15213, USA}
\affiliation{Department of Applied Physics, Yale University, New Haven, CT 06511, USA}
\author{Chao Zhou}
\affiliation{Department of Physics and Astronomy, University of Pittsburgh, Pittsburgh, PA 15213, USA}
\author{Chenxu Liu}
\thanks{Recently moved to: Pacific Northwest National Laboratory, Richland, WA 99354, USA}
\affiliation{Department of Physics, Virginia Tech, Blacksburg, Virginia 24061, USA}
\affiliation{Virginia Tech Center for Quantum Information Science and Engineering, Blacksburg, VA 24061, USA}
\author{Param Patel}
\affiliation{Department of Physics and Astronomy, University of Pittsburgh, Pittsburgh, PA 15213, USA}
\affiliation{Department of Applied Physics, Yale University, New Haven, CT 06511, USA}
\author{Xi Cao}
\author{Pinlei Lu}
\affiliation{Department of Physics and Astronomy, University of Pittsburgh, Pittsburgh, PA 15213, USA}
\author{Boris Mesits}
\affiliation{Department of Physics and Astronomy, University of Pittsburgh, Pittsburgh, PA 15213, USA}
\affiliation{Department of Applied Physics, Yale University, New Haven, CT 06511, USA}
\author{Maria Mucci}
\author{David Gorski}
\author{David Pekker}
\affiliation{Department of Physics and Astronomy, University of Pittsburgh, Pittsburgh, PA 15213, USA}
\author{Michael Hatridge}
\affiliation{Department of Physics and Astronomy, University of Pittsburgh, Pittsburgh, PA 15213, USA}
\affiliation{Department of Applied Physics, Yale University, New Haven, CT 06511, USA}
\affiliation{Yale Quantum Institute, Yale University, New Haven, CT 06511, USA}

\date{\today}

\begin{abstract}
    Increasing the fidelity of single-qubit gates requires a combination of faster pulses and increased qubit coherence. However, with resonant qubit drive via a capacitively coupled port, these two objectives are mutually contradictory, as higher qubit quality factor requires a weaker coupling, necessitating longer pulses for the same applied power. Increasing drive power, on the other hand, can heat the qubit's environment and degrade coherence. In this work, by using the inherent nonlinearity of the transmon qubit, we circumvent this issue by introducing a new parametric driving scheme to perform single-qubit control. We achieve rapid gate speed by pumping the transmon's native Kerr term at approximately one third of the qubit's resonant frequency. Given that transmons typically operate within a fairly narrow range of anharmonicity, this technique is applicable to all transmons. In both theory and experiment, we show that the Rabi rate of the process is proportional to applied drive amplitude cubed, allowing for rapid gate speed with only modest increases in applied power. In addition, we demonstrate that filtering can be used to protect the qubit's coherence while performing rapid gates, and present theoretical calculations indicating that decay due to multi-photon losses, even in very strongly coupled drive lines, will not limit qubit lifetime. We demonstrate single qubit gates as short as \unit[37.4]{ns} with fidelity as high as 99.91\%. We also present calculations indicating that this technique could reduce cryostat heating for fast gates, a vital requirement for large-scale quantum computers.  
\end{abstract}

\maketitle

\section{Introduction}
\label{msec:introduction}

High-fidelity single qubit control is one of the fundamental requirements for gate-based quantum computing.
While many factors can limit gate fidelity, such as cross-talk~\cite{sheldon2016procedure, mundada2019suppression} and leakage to non-computational states~\cite{motzoi2009simple}, the most fundamental are gate speed and qubit coherence~\cite{barends2014superconducting, kelly2014optimal,  sheldon2016characterizing, boixo2018characterizing, kjaergaard2020superconducting}, with recent improvements driven more by increased coherence than enhanced speed~\cite{place2021new, wang2021transmon}.
However, for superconducting qubits, single qubit gates typically use resonant driving of a qubit transition, in which the requirements for fast gates and coherent qubits are contradictory.

For instance, to protect the qubit coherence time, weakly coupled drive ports and heavily attenuated lines are typically used to suppress qubit photon leakage, thermal noise~\cite{wang2019cavity, wang2021transmon}, and out of band radiation~\cite{devoret2019pairbreaking}.
As qubit coherence increases, there is a trend towards more thorough and careful filtering. The increased losses of filtering and weaker qubit-drive line couplings together result in longer gate times at a given drive strength. However, achieving high gate fidelity requires we maintain or increase the ratio of qubit lifetime to gate time and so we attempt to compensate by increasing drive strength, which in turn increases heating of the cryostat. Heating of filter elements and drive lines~\cite{yeh2017microwave} can degrade the qubit by creating an excess thermal population at the qubit transition or higher frequencies (for example, at the superconducting gap) and/or heating the entire cryostat by exceeding the cooling capacity of the dilution refrigerator~\cite{krinner2019engineering}. 
Moreover, the thermal time constants (in the millisecond range) are much slower than the pulses (roughly 10s of nanoseconds range) or qubit coherence times ($\sim$\unit[100]{$\mu$s} to \unit[1]{ms}), and can produce effects which accumulate and persist across many experiments~\cite{yeh2017microwave}.
This situation is exacerbated by the necessity of scaling to larger quantum machines and thus increased qubit count and associated heating~\cite{boixo2018characterizing}.

To combat these limitations, better heat handling of each component, especially the attenuators on the control lines, has proven useful~\cite{yeh2017microwave, wang2019cavity}.
Another solution is to break the symmetry between input control drives and out-going qubit photon decay in the power domain using a non-linear filter~\cite{kono2020breaking}.
In this paper, we propose breaking the link between qubit coherence and driving by separating the two processes in frequency space, proposing and demonstrating a new single qubit control scheme based on parametric driving, which we term subharmonic driving.

In subharmonic driving we operate a transmon, a commonly used fixed-frequency qubit, by parametrically driving it at one third of its $\ket{g}\leftrightarrow\ket{e}$ transition frequency. 
Similar controlling methods are widely used in other systems, such as parametric amplification~\cite{clerk2010introduction, roy2016introduction} and parametric qubit-cavity~\cite{narla2016robust}, multi-cavity~\cite{sirois2015coherent, axline2018demand, zhou2023realizing} and multi-qubit gates~\cite{niskanen2007quantum, mckay2016universal}.
In each of these scenarios, far off-resonant terms in the Hamiltonian are utilized, consuming pump photons to produce effective lower-order interaction Hamiltonians which need not satisfy energy conservation.
The benefits of parametric driving include strong interactions and high on/off ratios; the use of reflective filters to protect mode frequencies while allowing strong driving is also well established~\cite{axline2016architecture, Rehammar2022LowpassFW, zhou2023realizing}.

The dominant nonlinear term in a transmon, the self-Kerr, is generated from the fourth order term of the Josephson junction's cosinusoidal potential, with magnitude typically ranging from \unit[-150 to -250]{MHz}~\cite{koch2007charge}. 
We pick the $4^{\text{th}}$ order term $\hat{q}^\dagger\hat{q}\hat{q}\hat{q} + h.c.$, where $\hat{q}^\dagger$ is the qubit creation operator, and parametrically drive the qubit at near one third of the qubit transition frequency.
Under these drive conditions, the chosen term annihilates three drive photons to create one qubit photon.
The effect of the drive is described by the effective Hamiltonian 
$\sim \eta^3\q^\dagger+h.c.$, where $\eta$ is the dimensionless amplitude of the coherent drive.
We note that the potential energy contains many other terms; however, these terms are suppressed as they are fast-rotating as can be seen by applying the rotating wave approximation (RWA).
We emphasize that the term we use is both ubiquitous and of very similar strength in every transmon qubit. This scheme can also be adapted to other qubits and systems by choosing an appropriate source of nonlinearity, for instance, $3^\text{rd}$ or $5^\text{th}$ order terms in qubits with asymmetric Josephson elements~\cite{noguchi2020fast}, and thus can be applied widely in superconducting circuits.

\begin{figure}[ht]
    \centering
    \includegraphics[width=0.9\linewidth]{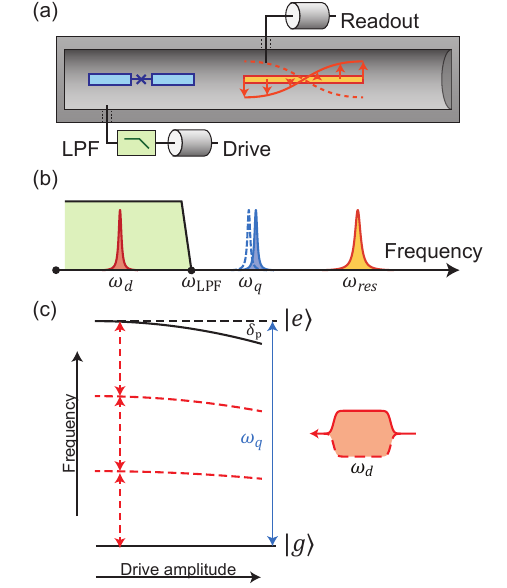}
    \caption{\textbf{Subharmonic driving schematic.}
    \textbf{a}  A transmon qubit (at left in blue) and a $\lambda/2$ resonator for readout (at right in orange) are supported on a sapphire chip inside an aluminum tube.  A low-pass filter (in green) is placed at the qubit drive port to suppress qubit photon leakage to the environment while allowing fast single qubit control at low frequencies. 
  \textbf{b} The frequency distribution of the subharmonic drive $\omega_d$~(red), qubit frequency $\omega_\text{ge}$~(blue), and the readout resonator frequency $\omega_{res}$~(orange). Only the subharmonic drive is within the pass-band of the LPF (green). The dashed, shifted peak of $\omega_\text{ge}$ represents the ac-Stark effect during the subharmonic drive.
    \textbf{c} The energy level diagram of subharmonic driving. The ac-Stark effect induced by subharmonic drive creates a detuning $\delta_p$ between the qubit's un-driven and driven frames. Flat-top pulses are used to control the qubit state. }
    \label{fig:schematic}
\end{figure}

The proposed control scheme has two main advantages.
The first advantage is that the drive frequency is separated from the qubit transition frequency. It allows us to suppress a primary relaxation channel in the system without affecting the ability to control the qubit by engineering the impedance at two widely separated frequencies.
To this end, we placed a reflective low-pass filter (LPF) with low absorption (S11($\omega_d$)=\unit[0.2]{dB}) and good isolation in the stop band (S21($\omega_q$)=\unit[61]{dB}) at the qubit drive port, as shown in Figure~\ref{fig:schematic}(a). This suppresses qubit photon leakage to the environment through resonant decay while allowing low-frequency drive photons to pass freely.
Therefore, the drive port can be more strongly coupled to the qubit for fast control without increasing the qubit's direct relaxation rate.  
The second advantage of subharmonic control is that the Rabi rate is proportional to the drive amplitude cubed because it is a three-photon driving process, which allows us to rapidly improve qubit gate speed with only a moderate increase of drive power.

In this paper, we experimentally demonstrate the concept of single qubit, subharmonic control and achieve gates as fast as \unit[37.4]{ns} with gate fidelities up to 99.91\% on a typical, unoptimized transmon qubit. We also present calculations which address two key questions about practical implementation of subharmonic gates.
First, we present a theoretical study of the effects of the low frequency lossy environment on the qubit's coherence, finding that this coupling offers a negligible decay channel even for very strong couplings.
Second, based on both theoretical calculations and measured parameters of our system, we find that the combination of low loss at base with realistic filter losses should allow high fidelity qubit gates with heating lower than conventional direct qubit drives.  

\section {Subharmonic driving theory}
\label{msec:theory}

The displaced Hamiltonian of a transmon under an off-resonant drive $\varepsilon(t)\text{cos}(\omega_d t)$ can be written as 
\begin{equation}\label{eq:Hamiltonian_displaced}
    \hat{H}^D/\hbar = (\omega_{q} - \alpha) \q^\dagger \q + \frac{\alpha}{12}(\q + \eta e^{-i\omega_d t} + h.c.)^4,
\end{equation}
where $\eta=i\beta(\omega_d)\varepsilon(t)$ is the dimensionless drive strength, and $\beta={2\omega_d}/[(\omega_q-\alpha)^2-\omega_d^2]$ (see Supplement Sec.~I~A).
Expanding the Hamiltonian $\hat{H}^D$, we find numerous $4^{th}$-order terms corresponding to different parametric processes of the transmon, which can be individually activated by driving at the correct frequencies.
For example, driving the term $4 \eta e^{i\omega_d t}\q\q\q +h.c.$ at $3(\omega_q+\alpha)$ activates the three-photon transition $\ket{g}\leftrightarrow\ket{h}$ (where $\ket{h}$ is the third excited state), while the two-photon transition $\ket{g}\leftrightarrow\ket{f}$, commonly seen in transmon spectroscopy (usually termed $gf/2$), can be activated by driving the term $6 (\eta^* e^{i\omega_d t})^2 \q\q +h.c.$ near $(2\omega_q +\alpha)/2$. 
For subharmonic driving, the term we are interested in is $4(\eta^* e^{i\omega_dt})^3 \q +h.c.$ Moving to the rotating frame at $3\omega_d \approx \omega_q$ and performing the RWA, we acquire the desired single qubit Rabi drive term, as well as a term proportional to $(\eta \eta^*)q^\dagger q$, which represents the ac-Stark effect during subharmonic drive.

In the end, the Hamiltonian has the form:
\begin{equation} \label{eq:Hamiltonian_RWA}
    \hat{H}^R_{\text{sub}}/\hbar = (2\alpha|\eta|^2-3\delta)\q^\dagger \q + \frac{1}{2}\alpha \q^\dagger \q^\dagger \q\q + \frac{1}{3}\alpha(\eta^3\q^\dagger+\eta^{*3}\q),
\end{equation}
with the drive detuning $\delta = \omega_d - \frac{1}{3}\omega_q$.  Equation~\eqref{eq:Hamiltonian_RWA} shows two important properties of subharmonic driving: the Rabi rate of the process $\Omega$ is proportional to $|\eta|^3$, and the ac-Stark shift $\Delta \omega$ during the subharmonic drive is proportional to $|\eta|^2$. 
The ac-Stark effect also adds a drive-dependent phase on the qubit when performing a gate, which needs to be considered and calibrated in all experiments, as the qubit frequency changes in response to the amplitude of the drive.

One potential concern about subharmonic driving is qubit photon decay through the drive line at low frequencies via multi-photon processes, which could limit coherence when we couple the transmon and drive line very strongly. We perform a detailed calculation in the Supplement Sec.~I~G and summarize key results here. For a transmon qubit coupled to a low-frequency lossy environment, the system-bath coupling strength is 
\begin{equation}
    \lambda(\nu)=\Theta(\nu)\frac{C_c}{\sqrt{C_r c}}\sqrt{\frac{\omega_q \nu}{2\pi v}},
    \label{eq:S-B coupling}
\end{equation}
where $C_{c}$ is the coupling capacitance between the qubit and the transmission line, $C_{r}$ is the capacitance of the qubit, $c$ is the characteristic capacitance of the transmission line, and $v$ is the speed of light in the transmission line~\cite{Blais2021}.
In our case, we put a cut-off (filter) function $\Theta(\nu)$ to suppress the high-frequency system-bath coupling. Specifically, the filter function is modeled as
\begin{align}
    \Theta(\nu) = \left\lbrace
    \begin{array}{cl}
        1 & \nu \leq \omega_q/3 + \vartheta \\
        0 & \nu > \omega_q/3 + \vartheta,
    \end{array}
    \right.
\end{align}
where $\omega_q/3+\vartheta$ is the bandwidth of the filter pass band. The decay rate through a three photon decay process can then be written as
\begin{equation}
    \Gamma_3 = \frac{243}{32 \pi^2} \frac{\gamma_1^3 \vert \alpha \vert^2}{\omega_q^4} \left( \frac{\vartheta}{\omega_q}\right)^2
    \label{eq:decay_rate}
\end{equation}
where $\gamma_1 = 2\pi \lambda^2$, $\lambda$ is the system-bath coupling strength. For a transmon system with typical properties, because $\alpha$ and $\gamma_1$ are both much smaller than $\omega_q$, the subharmonic decay rate $\Gamma_3$ is orders of magnitude smaller than the decay rate through resonant decay via internal losses or residual coupling to drive ports.
Therefore, $\Gamma_3$ can be safely neglected even for very strong couplings.

\section{experimental results}
\label{msec:experiment}

The subharmonic driving scheme was tested on a single qubit-resonator system, which had a transmon qubit and a $\lambda/2$ stripline resonator~\cite{wang2016schrodinger} with parameters commonly used in the field.
Specifically, a transmon qubit ($\omega_\text{ge}/2\pi=\unit[4.237]{GHz}$, $\alpha/2\pi=\unit[-148.7]{MHz}$, $T_1/T_{2R}/T_{2E}=\unit[41/21/72]{\mu s}$) and a $\omega_\text{res}/2\pi=\unit[6.498]{GHz}$ readout resonator were housed inside a \unit[4]{mm} diameter aluminum tube. The coupling rate of the resonator is $\kappa_\text{ext}/2\pi = \unit[1.44]{MHz}$ and the cross-Kerr between the qubit and the resonator is $\chi/2\pi=\unit[0.77]{MHz}$.
This system was chosen to represent a general transmon-cavity system without any special engineering or specific design requirements, demonstrating that the subharmonic driving scheme is widely applicable to transmon-based quantum processors.

\begin{figure}[ht]
    \centering
    \includegraphics[width=1\linewidth]{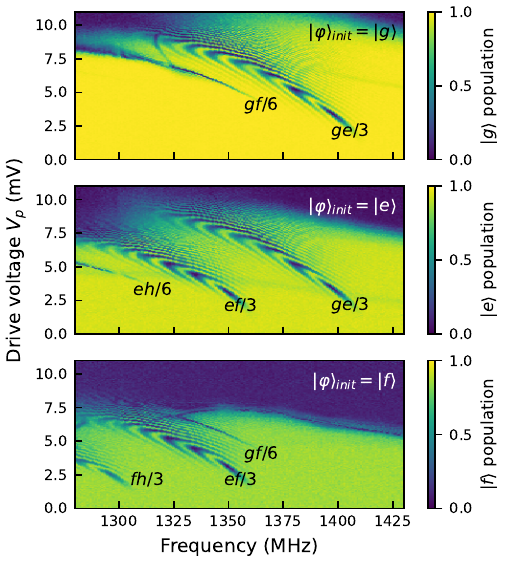}
    \caption{\textbf{Parametric processes near one third of the qubit frequency. }Multiple transitions can be activated by preparing initial states $\ket{g}$, $\ket{e}$, and $\ket{f}$ and sweeping drive amplitude and frequency. The drive voltage $V_p$ represents the peak amplitude of a pulse at the base stage of the dilution refrigerator.}
    \label{fig:power_rabi}
\end{figure}

To better understand subharmonic driving, we first performed Rabi experiments, sweeping the drive frequency and amplitude on the qubit to observe its population oscillating between two states.
Figure~\ref{fig:power_rabi} shows qubit response vs. drive frequency, amplitude, and initial state. 
The pulse used here was a 100 ns flat-top pulse with smoothed edges (see Supplement Sec.~I~B).
Preparing the qubit in the states $\ket{g}$, $\ket{e}$ and $\ket{f}$, we see the four-wave mixing $ge/3$, $ef/3$ and $fh/3$ processes, respectively, by activating the term $q^\dagger qqq + h.c.$. They are separated by a frequency of $\alpha/3$.
Furthermore, the 8-wave mixing processes $gf/6$ and $eh/6$ can be observed, with the term $q^\dagger q^\dagger q^6+h.c.$ activated. The $gf/6$ process is the subharmonic correspondence of the $gf/2$ peak that is often seen in qubit spectroscopy.

We also observe a state-dependent upper boundary above which the coherent oscillations disappear and the transmon population moves to higher excited states. Such a boundary puts a maximum speed limit of our subharmonic gates, but is beyond the scope of this paper and the subject of a separate investigation\cite{xia2025exceeding}. Within this upper boundary, all of these parametric processes can be used to perform qubit controls without appreciable degradation of the qubit's coherence. In this work, we focus on the $ge/3$ process and use it to build fast single-qubit gates.

\begin{figure}[ht]
    \centering
    \includegraphics[width=1\linewidth]{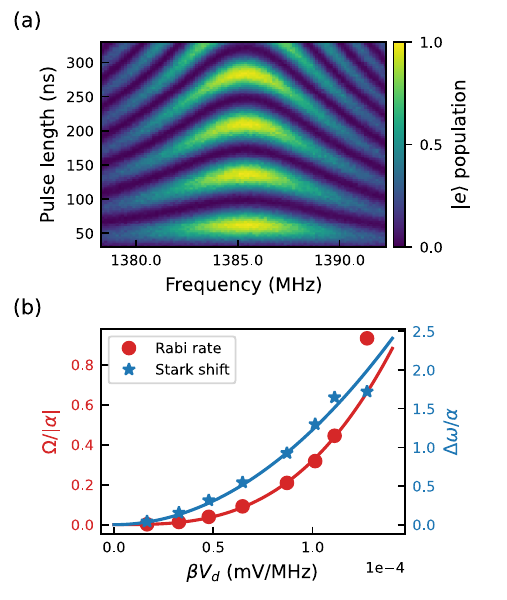}
    \caption{
    \textbf{Rabi rate and ac-Stark shift vs. subharmonic driving strength.}
    \textbf{a} A Rabi experiment for subharmonic driving of the qubit as a function of pulse duration and drive frequency at fixed drive amplitude of $\unit[4.71]{mV}$. At this amplitude, the Rabi rate is \unit[$2\pi\times$13.55]{MHz} and the drive-induced qubit ac-Stark shift is \unit[$2\pi\times$-81.25]{MHz}, equivalent to 0.55 times of the transmon anharmonicity $\alpha$.
    \textbf{b} Rabi rate and qubit ac-Stark shift vs. drive amplitude. The Rabi rate (red squares) is proportional to drive amplitude cubed. The ac-Stark shift (blue stars) is proportional to drive amplitude squared. A single fit function describes both sets of data with only one free parameter, the scaling factor $k$ converting pulse amplitude $\beta V_p$ to the effective driving strength $\eta$.
    }
    \label{fig:Rabi}
\end{figure}

Our next step is driving the qubit with fixed power (and hence ac-Stark shift) and variable pulse duration at a given pulse amplitude to find the Rabi rate and frequency. As shown in Figure~\ref{fig:Rabi}(a), the subharmonic Rabi oscillation versus time shows a chevron pattern similar to the pattern observed using a resonant drive. By fitting the data, we extract both the ac-Stark shift induced by the off-resonant drive on the qubit and the Rabi rate at the given drive strength.

The experiment was then repeated with other pulse amplitudes. 
The drive amplitude in millivolts is the pulse peak amplitude at the base stage of the dilution refrigerator. The microwave configuration on the input line is shown in Extended Data Figure~\ref{fig:diagram}. The method of calibrating the pulse amplitude in the base stage is discussed in the Supplemental Sec.~II~B.
After calibration, we assume that the voltage is linearly proportional to the drive strength $\eta$ on the qubit.
As shown in Figure~\ref{fig:Rabi}(b), with a drive amplitude of \unit[1.19]{mV}, the slowest Rabi rate was only \unit[$2\pi\times$0.27]{MHz}. When the drive amplitude was increased by 7.1 times to \unit[8.48]{mV}, the Rabi rate reached \unit[$2\pi\times$68.04]{MHz}, which increased by around 250 times. The qubit also experiences an ac-Stark shift around 1.6 times of its anharmonicity. 
The Rabi rate $\Omega(V_d)$ and the ac-Stark shift $\Delta\omega(V_d)$  as a function of drive strength $V_d$ are fitted simultaneously with a single free parameter $k$, which relates the room temperature and cryogenic drive voltages:
\begin{align}
    \left\lbrace
    \begin{array}{cl}
        \Delta\omega(V_d) &= 2\alpha (k\beta V_d)^2\\
        \Omega(V_d) &= \frac{2}{3}\alpha (k\beta V_d)^3,
    \end{array}
    \right.
\end{align}
where the transmon self-Kerr $\alpha$ and $\beta$ are known parameters measured using qubit spectroscopy. In Figure~\ref{fig:Rabi}(b), the ac-Stark shift is normalized by the transmon anharmonicity $\alpha$ and the data fit well with the fitting function, with $k=\unit[2\pi\times1.25\times10^3]{MHz/mV}$.
The data point with the highest drive amplitude is plotted but not used in the fitting.  In general, the details of the qubit response at the very strongest drives, here with a Rabi rate approaching the qubit anharmonicity, is subject to a number of complicating factors and requires further investigation in future work.

These Rabi experiments demonstrate subharmonic driving can perform fast controls on quite standard transmon qubits. To use this scheme practically, we also need to develop a procedure for calibrating a high-fidelity single-qubit gate with well-defined parameters.
For on-resonance driving, both deterministic tune-up~\cite{ReedThesis} and randomized benchmarking procedures have been well developed~\cite{kelly2014optimal}.
However, these procedures cannot be directly applied to subharmonic gates tune-up without first addressing the issue of drive-induced frequency shift.
When the generator frame and the qubit frame are rotating at different speeds, phase tracking is required.
In the case of subharmonic driving, as shown in Eq.~\eqref{eq:Hamiltonian_RWA} and Figure~\ref{fig:Rabi}, the qubit frequency is shifted. Therefore, the qubit frame rotates at a different speed relative to the generator frame during the drive and effectively adds a phase shift $\varphi_i$ to the qubit. Such phase shift can be calculated as
\begin{equation}
    \varphi_i = \int_{t_i}^{t_{i+1}}\omega_q+\Delta\omega(t)-3\omega_d(t)dt,
    \label{eq:gate_phase}
\end{equation} 
where $t_i$ and $t_{i+1}$ are the starting time and ending times of gate \textit{i}.
One way to correct the phase shift is by applying a virtual-\textit{Z} gate after the pulse\cite{mckay2017efficient}. Alternately, we can apply a time-varying frequency modulation to the pulse, which changes the instantaneous frequency of the pulse based on the qubit's ac-Stark shift. Ideally, a frequency-modulated pulse can always satisfy $3\omega_d(t)=\omega_q+\Delta\omega(t)$, which makes phase tracking unnecessary. However, in experiments, we found that phase tracking is usually still required to remove residual errors over many pulses.

\begin{figure}[ht]
    \centering
    \includegraphics[width=0.9\linewidth]{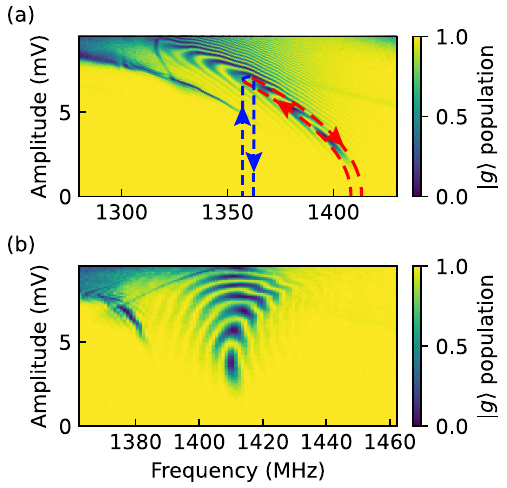}
    \caption{
    \textbf{Amplitude-dependent frequency modulation of the drive frequency.} \textbf{a} With the drive frequency following the ac-Stark shift of the transmon (red trace), a frequency chirped pulse is always resonant with the subharmonic process. Compared with fixed-frequency drive (blue trace), it avoids directly crossing other processes. \textbf{b} A power Rabi experiment with frequency chirping. The frequency axis here is the initial frequency of the pulse.}
    \label{fig:power_rabi_chirp}
\end{figure}

Additionally, applying a fixed-frequency pulse with large ac-Stark shifts can cause direct excitation to non-computational states.
For example, as shown in Fig.~\ref{fig:power_rabi_chirp}a, when the qubit is prepared in the ground state and driven at fixed frequency $\omega_d = \omega_{q}/3 + \delta$, the $gf/6$ process can be activated during the ramp up/down period of a pulse if the detune $\delta$ is larger than one sixth of the transmon anharmonicity  $\alpha$. 
By following the transmon frequency during a drive, chirped pulses avoid directly crossing and activating unwanted processes, and thus reduce leakage.
Figure~\ref{fig:power_rabi_chirp}(b) shows the Rabi experiment with frequency modulation using a quadratic model $\Delta\omega=2\alpha|\eta|^2$, as predicted by the $4^{th}$-order truncated Hamiltonian.  The ac-Start shift of the transmon is mostly removed.
The remaining distortion of the Rabi fringes arises primarily from imperfect pulse envelopes and the mismatch of instantaneous frequency and drive amplitude.  We can reduce the distortion by performing pulse envelope correction and including higher-order Hamiltonian terms to obtain a more accurate modulation model.

To benchmark our subharmonic single qubit gates, we choose the usual $\pi$ and $\pi/2$ rotation over the \textit{X}, \textit{Y}, and \textit{Z} axes as the basic operations. Rotation about the \textit{Z} axis is realized through \textit{VZ} gate, which has no cost and fidelity close to one\cite{mckay2017efficient}. For the $X$ and $Y$ gates, the drive amplitude needs to be balanced between errors due to qubit decay, leakage to non-computational states, hardware limitations, and other factors.
Because of control hardware limitations, we found best results when performing rotations of $\pi$ and $\pi/2$ with \unit[50.9]{ns} and \unit[37.4]{ns} long smoothed flat-top pulses. The details of a gate sequence-based calibration procedure that calibrates each parameter are explained in Supplement Sec.~II~C. 

Randomized benchmarking and interleaved randomized benchmarking \cite{knill2008randomized, magesan2012characterizing, magesan2012efficient} are used to calibrate the average fidelity of all Clifford gates and the fidelity of specific gates, including $X$, $\sqrt{X}$, $Y$ and $\sqrt{Y}$ gates.
The fidelity of the interleaved gate is calculated using~\cite{magesan2012efficient},
\begin{equation}
    1-F = \frac{1-p_\text{gate}/p_\text{ref}}{2}.
    \label{eq:rb_fidelity}
\end{equation}

\begin{figure}[ht]
    \centering
    \includegraphics[width=0.9\linewidth]{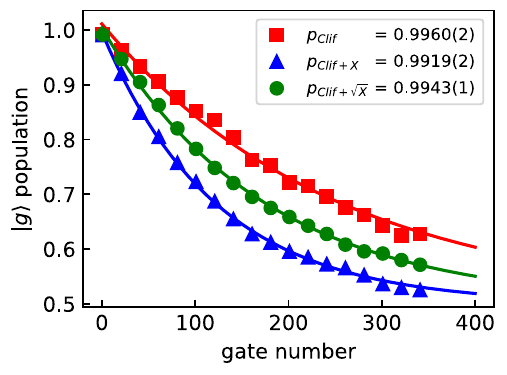}
    \caption{
    \textbf{Randomized gate benchmarking}. Gate fidelities
    are calibrated with randomized and interleaved randomized benchmarking. The red trace shows the average fidelity of Clifford gates using subharmonic driving. The green and blue traces show the average fidelity of random sequences with a specific interleaved Clifford gate ($X$ and $\sqrt{X}$ here), respectively, and are used to extract the fidelities of individual gates.}
    \label{fig:RB}
\end{figure}

As shown in Figure \ref{fig:RB}, the average Clifford gate fidelity is 99.604(9)\%, and the gate fidelities of $X$, $\sqrt{X}$, $Y$, and $\sqrt{Y}$ are 99.79(1)\%, 99.91(1)\%, 99.76(2)\%, 99.91(1)\% respectively. To maintain this fidelity, a stable environment for room-temperature control hardware is crucial. The subharmonic gates are three times more sensitive to the amplitude and phase drift than resonant gates, because of the Rabi rate's cubic dependence on drive amplitude. To reduce pulse parameter drifting, we stabilized the room-temperature equipment's environment by sealing it in a PID controlled temperature stabilizing box, which is discussed in detail in Methods and Supplement Sec.~II~D.

\section{Potential for Reduced Heating and fast subharmonic gates}
\label{msec:heating}
We have shown experimentally that a subharmonic drive can perform fast and high-fidelity single qubit control, and found theoretically that strong couplings to low-frequencies are not associated with enhanced qubit losses.
Of far greater concern is the behavior of the low-pass filter, which protects the qubit. The filter's figures of merit are the absorptive losses in reflection and insertion loss at the qubit's transition frequency (which protect the qubit from the environment noise and Purcell decay into the filter) and the losses in transmission at the drive frequency (which contribute to cryostat heating).

Measuring the filter used in this experiment (Mini-circuits ZLSS-A2R8G-S+) at room temperature, the impedance is $\unit[6.66-123.3i]{\Omega}$ at the qubit $g-e$ frequency. Due to this impedance mismatch, the decay rate through the qubit port at $\omega_q$ is improved from $\kappa(\omega_q)/2\pi = \unit[0.55]{kHz}$ to $\kappa_\text{LPF}(\omega_q)/2\pi = \unit[18.2]{Hz}$ using the filter. The subharmonic decay rate $\kappa_{sub}$ is much smaller than $\kappa_{LPF}$ and can be neglected.  To verify the performance of the LPF, we measured three qubits with a LPF on the qubit drive port and one qubit without the LPF. Although the improvement of qubit coherence time is not as good as theory, we still see a clear improvement (see Supplement Sec.~II~A).  The discrepancy is likely due to the loss and electrical length of the cable between the qubit and the filter. An on-chip filter\cite{} can further reduce photon loss and improve filter protection.

Subharmonic driving also offers new possibilities to address cryostat and component heating. It is a common practice to attenuate the input signal by \unit[60]{dB} or higher to reduce thermal noise to a very small residual photon occupancy in the drive lines at the readout resonator and qubit frequencies~\cite{wang2022towards, chakram2021seamless}.

\begin{figure}[ht]
    \centering
    \includegraphics{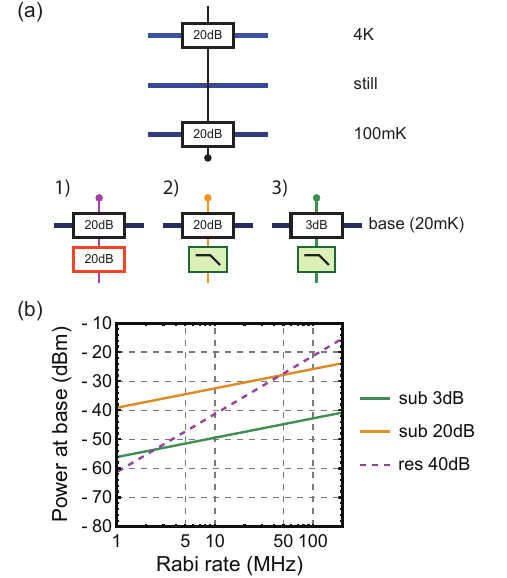}
    \caption{\textbf{Resonant drive vs. subharmonic drive.}
    \textbf{a} Different configurations of input line for regular drive and subharmonic drive. More attenuation is required to better protect the qubit from thermal noise in resonant drive method, while attenuators can be removed and replaced with reflective filters in subharmonic drive method. A commercial LPF is used on the subharmonic drive input line.
    \textbf{b} A comparison of dissipated heat vs. Rabi rate $\Omega$ between different input line configurations that have a $T_1$ limit of around \unit[1]{ms}. The regular drive is shown in dashed line, and subharmonic drive is shown in solid line.}
    \label{fig:heat_comparison}
\end{figure}

These attenuators put a heat load on the refrigerator and can become a limit for near-term quantum machines.
Figure~\ref{fig:heat_comparison} shows an estimate of the heat dissipated at the base stage of a dilution refrigerator in various driving configurations. To give a fair comparison, the port coupling strengths limit the qubit lifetime to $T_1=\unit[1]{ms}$ based on finite element circuit modeling and the measured impedance of our current filter. Using resonant driving and configuration 1 shown in Figure~\ref{fig:heat_comparison}(a) (a high-attenuation conventional drive line), approximately \unit[-30]{dBm} of power is dissipated at the base stage to achieve a $\pi$-pulse time of \unit[10]{ns}.
The current limited cooling capacity of our dilution refrigerator ($\sim \unit[10]{\mu W}$ at $\unit[20]{mK}$) allows only tens of single-qubit drives in parallel before significantly heating the base stage.
Moreover, the attenuator itself as a dissipative element can be heated and generate thermal photon in turn. Commercial attenuators have been shown to cool much slower than the typical experiment repetition rates, creating further complications~\cite{yeh2017microwave}. It would be advantageous to simply remove these components.

As shown in Fig.~\ref{fig:heat_comparison}, we compare conventional drive lines with \unit[20]{dB}  base attenuation and a LPF (configuration 2, close to our experimental configuration) and LPF only at base (configuration 3) for subharmonic gates.
The third configuration removes attenuators from the base plate completely and alleviates the heat dissipation at the base stage. The remaining \unit[3]{dB} attenuation represents our estimate of realistic losses on components other than RF attenuators, such as Eccosorb filters and coaxial cables, which should dominate over the insertion loss of the LPF. Reducing the losses at base allows us to take full advantage of subharmonic driving; at a Rabi rate of \unit[50]{MHz}, heat dissipation for configuration 3 is improved by around \unit[20]{dB} relative to the conventional drive configuration, allowing 100 times more qubits to be controlled for the same drive power and realistic assumptions about the drive line configuration.

\section{Conclusion}
\label{msec:conclusion}
In conclusion, we have shown that subharmonic driving is a new single-qubit control scheme with fast gate speed and high fidelity. By breaking the symmetry between drive and decay of a qubit in the frequency domain, we protect the qubit from resonant relaxation and control the qubit without any need for resonant access.
In addition, the cubic dependence of the gate speed on the drive strength could reduce the heat dissipated during qubit driving for appropriately designed reflective drive lines.

In the experiment single-qubit gates with fidelity of 99.9\% are realized. However, higher gate fidelity can be achieved. The gate time currently used to achieve 99.9\% fidelity is much slower than the maximum Rabi rate that we measured in this experiment. This is because of the pulse distortion and imprecise control of FPGA hardware, we believe gate times of \unit[10]{ns} or lower are possible to achieve.
In addition, DRAG correction or other pulse engineering techniques\cite{motzoi2009simple, khaneja2005optimal, caneva2011chopped} can be implemented to reduce gate error and improve gate speed. With moderate improvement in qubit coherence times and gate time of \unit[10]{ns}, gate fidelities of 99.99\% or higher are achievable, and, together with reduced cryostat heating, make subharmonic gates a compelling method for single-qubit control in future large-scale quantum computers.

\section{Acknowledgments}
This research, including funding for M.X., C.Z. and C.L. was supported by the U.S. Department of Energy, Office of Science, National Quantum Information Science Research Centers, the Co-design Center for Quantum Advantage (C2QA) under Contract No. DE-SC0012704.  Support was also provided by the Air Force Office of Strategic Research under award FA9550-15-1-0015, and the National Science Foundation PIRE HYBRID program under contract 1743717.  

\section{Author contributions}
M.J.H. conceived and supervised the project with help from D.P. on theoretical analysis.  M.X., C.Z., X.C., P.L., and M.M. acquired data for the project; P.L. contributed important software tools for randomized benchmarking, while B.M. built and tested with M.X. temperature stabilization hardware for the control electronics. P.J.P. designed and fabricated the samples.  C.Z. and D.G. contributed to theoretical calculations of the subharmonic gate, while C.L. performed the study of multi-photon decay channels.  M.X. led final data acquisition and the gate speed/heating analysis. M.X. and M.J.H. drafted the manuscript, to which all authors contributed. 

\section{Competing Interests}
M.J.H. receives consulting fees and/or holds equity in Quantum Circuits, Inc.

\bibliography{refs}

\newpage

\renewcommand\figurename{Extended Data Figure}
\renewcommand\thefigure{1}
\section{Methods}
\subsection{Experiment setup}
\label{msec:experiment_setup}
The metal block housing the qubit and cavity's tube is made of Al-6061 alloy with a \unit[4]{mm} diameter hollow tube and three coupling ports. One of the ports is coupled to the transmon for single qubit control. The other two ports are coupled to the resonator for the dispersive readout
. In this experiment, only one of the readout ports is connected with the output coaxial line and the other one is removed. The transmon qubit and the $\lambda/2$ stripline resonator are fabricated on a \unit[421]{nm} thick sapphire substrate with around \unit[200]{nm} Ta film except for the Josephson junction of the transmon, which is a Al/AlO$_\textrm{x}$/Al junction. 

The experiment setup is shown in the Extended Data Figure 1. We used a low-pass filter, Mini-circuits ZLSS-A2R8G-S+, to protect the qubit and perform subharmonic drives. The sample is mounted on the base plate of the dilution refrigerator, which is operated at around \unit[20]{mK}. The qubit driving pulses and readout pulses are generated by direct digital synthesis (DDS) using the Xilinx RFSoC ZCU216 based on the QICK~\cite{stefanazzi2022qick}. For the qubit drive, before the signal is sent to the sample, a small fraction is split using a directional coupler, then sent back to the digitizer for monitoring pulse stability.

\subsection{Temperature stabilization}
\label{msec:temperature}
The performance of room temperature electronics depends on their operating status and environment. 
For example, the gain of a power amplifier changes as environment temperature changes. In monitoring our pulse properties, we found the operating temperature, which is regularly perturbed by air-conditioning cycles in the room air, has the most significant effect on our system performance.
A clear dependence on devices' temperature can be observed, As shown in the supplement Sec.~II~D. While a few percents of relative variation in power and phase can often be ignored in simple one or two pulse protocols, such as $T_1$ or $T_2$ measurements, it is a limiting factor for high-fidelity quantum control, and it is important for building a stable quantum computer without frequent pulse calibration. 
Also, this is especially important for subharmonic driving, since it is three times more sensitive to the fluctuation of the pulses' magnitude and phase.

\begin{figure*}[ht]
    \centering\includegraphics{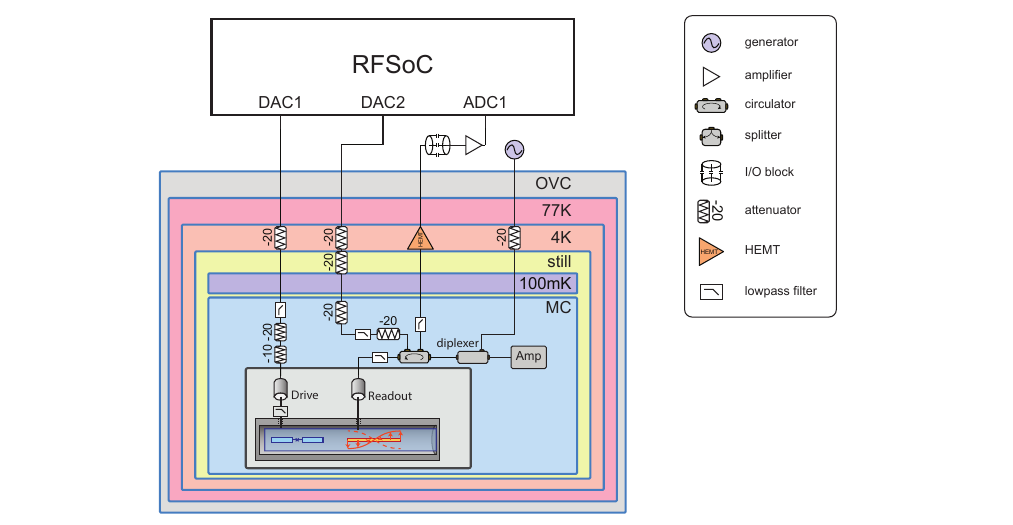}
    \caption{Cryogenic and measurement diagram of the sub-harmonic qubit drive experiment. In the qubit drive setup, a small part of the drive power 
 for monitoring drive stability are directly sent back to the digitizer using a \unit[10]{dB} directional coupler. All AWGs, digitizers and generators are syncronized to a Rubidium frequency standard. }
    \label{fig:diagram}
\end{figure*}
\end{document}


\preprint{APS/123-QED}

\title{Fast superconducting qubit control with subharmonic drives: Supplementary information}


\author{Mingkang Xia}
\affiliation{Department of Physics and Astronomy, University of Pittsburgh, Pittsburgh, PA 15213, USA}
\affiliation{Department of Applied Physics, Yale University, New Haven, CT 06511, USA}
\author{Chao Zhou}
\affiliation{Department of Physics and Astronomy, University of Pittsburgh, Pittsburgh, PA 15213, USA}
\author{Chenxu Liu}
\thanks{Recently moved to: Pacific Northwest National Laboratory, Richland, WA 99354, USA}
\affiliation{Department of Physics, Virginia Tech, Blacksburg, Virginia 24061, USA}
\affiliation{Virginia Tech Center for Quantum Information Science and Engineering, Blacksburg, VA 24061, USA}
\author{Param Patel}
\affiliation{Department of Physics and Astronomy, University of Pittsburgh, Pittsburgh, PA 15213, USA}
\affiliation{Department of Applied Physics, Yale University, New Haven, CT 06511, USA}
\author{Xi Cao}
\author{Pinlei Lu}
\affiliation{Department of Physics and Astronomy, University of Pittsburgh, Pittsburgh, PA 15213, USA}
\author{Boris Mesits}
\affiliation{Department of Physics and Astronomy, University of Pittsburgh, Pittsburgh, PA 15213, USA}
\affiliation{Department of Applied Physics, Yale University, New Haven, CT 06511, USA}
\author{Maria Mucci}
\author{David Gorski}
\author{David Pekker}
\affiliation{Department of Physics and Astronomy, University of Pittsburgh, Pittsburgh, PA 15213, USA}
\author{Michael Hatridge}
\affiliation{Department of Physics and Astronomy, University of Pittsburgh, Pittsburgh, PA 15213, USA}
\affiliation{Department of Applied Physics, Yale University, New Haven, CT 06511, USA}
\affiliation{Yale Quantum Institute, Yale University, New Haven, CT 06511, USA}

\date{\today}
             
\maketitle

\section{Theory}

\subsection{Transmon under subharmonic drive}\label{ssec:subharmonic_drive}
The un-driven transmon Hamiltonian can be written as

\begin{equation}
    \begin{split}
        \hat{H}_0&=\sqrt{8E_CE_J}\q^\dagger \q - \frac{E_C}{12}(\q^\dagger+\q)^4\\
        &=\hbar(\omega_q-\alpha)\q^\dagger \q+\hbar \frac{\alpha}{12}(\q^\dagger+\q)^4,
    \end{split}
    \label{eq:transmon_h}
\end{equation}
where $E_J$ and $E_c$ are the Josephson energy and charging energy of the transmon qubit, respectively, $\alpha = -E_c$ is the anharmonicity of the qubit, $\hbar\omega_q = \sqrt{8E_c E_J} -E_c$ is the qubit angular frequency, and $\hat{q}$ is the photon annihilation operator of the qubit. In the expansion of $(\q^\dagger+\q)^4$, most $4^\text{th}$ order terms are fast oscillating in the absence of drives and can be neglected under the rotating wave approximation (RWA). However, these fast-oscillating terms are important for parametric driving, and so we cannot immediately apply the rotating wave approximation. The Hamiltonian of a transmon under a single microwave drive at frequency $\omega_d$ and amplitude $\varepsilon$ can be written as
\begin{equation}
    \begin{split}
        \hat{H} &= \hat{H}_0 + \hat{H}_\text{drive}\\
        &=\hbar(\omega_q-\alpha)\q^\dagger \q+\hbar \frac{\alpha}{12}(\q^\dagger+\q)^4 + i\hbar \epsilon(t)(\q^\dagger - \q),
    \end{split}
\end{equation}
where
\begin{equation}
    \epsilon(t)=
    \begin{cases}
        \vep(t)e^{-i\omega_d t}+\vep^*(t)e^{i\omega_d t} & 0<t<t_\text{gate}\\
        0 & \text{otherwise}
    \end{cases}.
\end{equation}
 
If the drive frequency is nearly on-resonant with the qubit's transitions, this equation will yield the typical directly driven transmon dynamics.  On the other hand, for a far-detuned drive, it is useful to continue by applying a  displacement transformation $\hat{D}(t) = e^{z\q^\dagger - z^*\q}$ which cancels the driving term, moving its effects into the qubit operator. If we define $z$ as
\begin{align}
    z &= -\frac{i\vep(t)}{\omega_d - \omega_q^{'}}e^{-i\omega_d t} + \frac{i\vep^*(t)}{\omega_d + \omega_q^{'}}e^{i\omega_d t} & \text{and} & & \omega_q^{'} &= \omega -  \alpha,
\end{align}
the driving term is canceled up to a scalar $C$.
After the displacement transformation, the Hamiltonian can be written as
\begin{equation}
    \begin{split}
        \hat{H}^D/\hbar 
        &= \hat{D}\hat{H}\hat{D}^\dagger/\hbar + i\dot{\hat{D}}\hat{D}^\dagger\\
        &= \omega_q^{'} \q^\dagger \q + \frac{\alpha}{12}(\q^\dagger + \q -z^* -z)^4 + C,\\
    \end{split}
\end{equation}
which can be further simplified to:
\begin{equation}
    \hat{H}^D/\hbar = \omega_q^{'} \q^\dagger \q + \frac{\alpha}{12}(\q^\dagger + \q + \eta e^{-i\omega_d t} + \eta^* e^{i\omega_d t})^4 \textrm{, with } \eta = \frac{2i\omega_d\vep(t)}{\omega_q^{'2} - \omega_d^2}
\label{eq:SI_HD}
\end{equation}

The fourth order component of Eq.~\ref{eq:SI_HD} conceals many potential dynamical behaviors of the qubit that can be activated by applying the drive at specific  frequencies $\omega_d$~\cite{ReedThesisSM}. Generally, each term in the expansion of this fourth-order component involves 1-3 qubit operators $\hat{q}$ or $\hat{q}^\dagger$, and, correspondingly 3-1 `pump' waves $\eta$ or $\eta^*$. For our purposes, we are interested in the terms that activate the single-photon excitation ($\ket{g} \leftrightarrow \ket{e} $ transition) on the qubit, which therefore should contain one qubit operator and three pump waves.  Considering the drive we applied is at around one-third of the qubit frequency, i.e. $\omega_d = \omega_q/3+\delta$, we move to the rotating frame that rotates at $3\omega_d$ and observe the effect of this pump:
\begin{equation}
    \hat{H}^R/\hbar = \hat{R} \hat{H}^D \hat{R}^\dagger/\hbar + i\dot{\hat{R}}\hat{R}^\dagger \textrm{, with } \hat{R} = e^{3i\omega_d t \q^\dagger \q}.
\end{equation}
After t+his transformation, the first term in $\hat{H}^D$ yields:
\begin{equation}
    \hat{R} (\omega_q^{'}\q^\dagger \q) \hat{R}^\dagger+ i\dot{\hat{R}}\hat{R}^\dagger = (-\alpha - 3\delta)\q^\dagger \q.
\end{equation}
The second term in $\hat{H}^D$ after the transformation $\hat{R}$ can be expanded and each individual term has the general form of $c_{jkmn} q^{\dagger j} q^{k} e^{-i(m-n+3j-3k)\omega_d t}$, with $j,k,m,n$ being non-negative integers and $j+k+m+n=4$. The terms that satisfy $m-n+3j-3k=0$ survive under RWA and have a significant effect on qubit evolution. In summary, the Hamiltonian has the form:
\begin{equation}
    \begin{split}
        \hat{H}^R/\hbar &= (\hat{H}_\text{Stark} + \hat{H}_\text{Kerr} + \hat{H}_{d}^{eff})/\hbar\\
        &= (2\alpha|\eta|^2-3\delta)q^\dagger q + \frac{\alpha}{2}q^\dagger q^\dagger q q + \frac{\alpha}{3}(\eta^3 q^\dagger + \eta^{*3} q).
    \end{split} 
    \label{eq:SH_hamiltonian}
\end{equation}

For a given driving amplitude $\vep(t)$, we can choose a drive detuning $\delta$, which satisfies $2\alpha|\eta|^2-3\delta = 0$, such that the first term in Eq.~\ref{eq:SH_hamiltonian} is equal to 0. In the frame rotating at the frequency $3\omega_d = \omega_q+2\alpha|\eta|^2$, the Hamiltonian is
\begin{equation}
    \begin{split}
        \hat{H}^R/\hbar &=  \frac{\alpha}{2}q^\dagger q^\dagger q q + \frac{\alpha}{3}(\eta^3 q^\dagger + \eta^{*3} q),
    \end{split} 
    \label{eq:SH_hamiltonian_2}
\end{equation}
which has exactly the same form as a transmon under a $\ket{g} \leftrightarrow \ket{e} $ on-resonance drive~\cite{blais2021circuitSM}, with the drive strength$\frac{\alpha}{3}\eta^3$.
When $6^{th}$ order and higher order terms are included, we can follow the same steps and obtain a more accurate model of the ac-Stark shift and Rabi rate, and investigate higher-order parametric processes.

\subsection{Controlling pulses}
\label{ssec:flat-top}

Defining the controlling pulse as
\begin{equation}
    \label{eq:general_pulse}
    \varepsilon(t)=A(t)\text{cos}(\omega t+\varphi(t)),
\end{equation}
a pulse can be modulated in amplitude and frequency.

The amplitude modulation $A(t)$ of a pulse decides its shape. One of the most simple and commonly used pulse shape for transmon control is the Gaussian pulse, which has the benefit of having no side peaks in frequency domain and a relatively narrow frequency distribution. 

Another type of frequently used pulse shape is the flat-top pulse, or the smoothed square pulse. Flat-top pulses were used in this experiment for the qubit control.
Square pulses make full use of the drive under a given drive length and maximum amplitude. To reduce the side peaks in the frequency domain, the edges of the pulses are often smoothed. The definition of flat-top pulses can be different depending on the definition of the ramp-up/down.
The flat-top pulses used in the experiments are defined with $\tanh$ functions. The flat-top pulse starting at $t=0$ and length of $t_0$ is defined as:
\begin{equation}
    \label{eq:flat-top}
    A(t) = 
    \begin{cases}
        \frac{1}{2}A_0(\tanh(kt-t_1) - \tanh(k(t-t_0)+t_1)) & 0<t<t_0\\
        0 & \text{otherwise}
    \end{cases}
\end{equation}
where $k$ defines the ramping speed and $t_1/k$ defines the mid-points of the pulse ramp.
\begin{figure}[ht]
    \centering
    \includegraphics{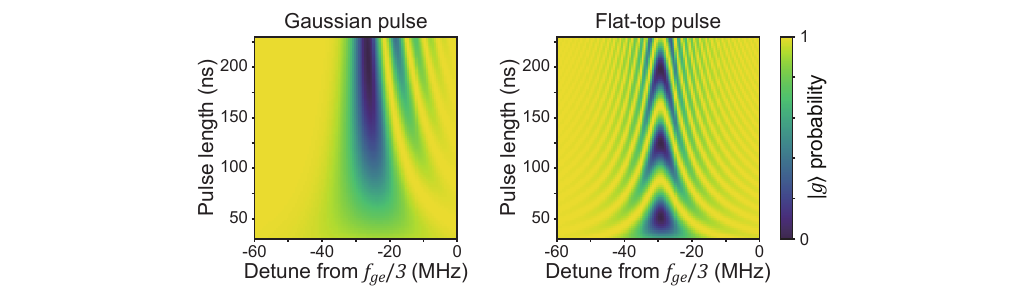}
    \caption{A simulation of fixed frequency Rabi experiment with the Gaussian pulse and the flat-top pulse.  }
    \label{fig:pulse_comparison}
\end{figure}
 
Figure~\ref{fig:pulse_comparison} shows the numerical simulation of the subharmonic Rabi process with fixed driving amplitude $A_0$ and varying length using Gaussian pulses and flat-top pulses.
As a result of the AC-Stark shift, the Rabi fringes of Gaussian pulses are distorted, and it takes a much longer time to perform $\pi$ and $\pi/2$ rotations. The result of the flat-top pulse shows a pattern similar to the resonantly driven Rabi experiment and is the same as experiment result in Fig.~\ref{fig:time_rabi_all}. 

The frequency modulation $\omega_{\text{mod}}(t) = d\varphi(t)/dt$ creates a pulse with time-varying frequency, which is useful to deal with phase tracking and the mode's AC-Stark shift in this experiment. To track the evolution of the qubit under drives, the instantaneous frequency of the modulation is one-third of the frequency shift from the transmon $g-e$ transition, i.e. $\omega_{\text{mod}}(t)=\Delta\omega_{\text{Stark}}(t)/3$.
Also, flat-top pulses have the advantage of facilitating the calibration of the added-phase resulting from the off-resonance drive, which is crucial for achieving phase-coherent qubit control and will be discussed in the next section.

\subsection{Phase correction of subharmonic gate}
\label{ssec:Phase_correction}
Because of the Stark effect during the subharmonic drive, performing a pulse will effectively add a phase to the qubit. Such phase drift needs to be tracked and fixed by applying phase corrections to all of the following pulses to rotate the qubit in the desired direction. It can be considered as adding a virtual-Z gate\cite{mckay2017efficientSM} between two subharmonic gates. Phase tracking is also often required in other kinds of superconducting qubit control, such as the parametric two-qubit gate and the CR gate. 

In the case of subharmonic driving, the phase difference between the drive frame and the qubit frame can be written as
\begin{equation}
    \vph(t_n) = \sum_{i=0}^n\vph_i = \varphi_0+\sum_{i=1}^n\int^{t_{i}}_{t_{i-1}} \omega_{q}(t) - 3\omega_{d}(t) dt
    \label{eq:Phase}
\end{equation}
where $t_i$ is the start time of pulse \emph{i} in the whole sequence.

\begin{figure}[ht]
    \centering
    \includegraphics{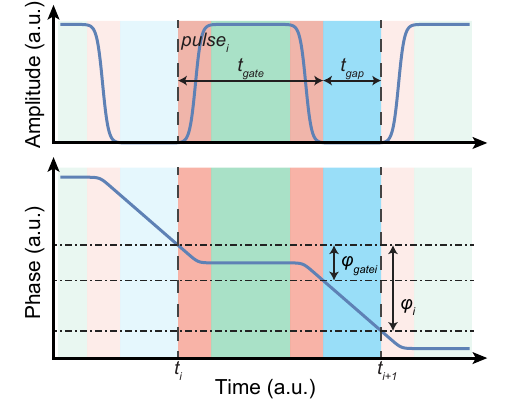}
    \caption{A pulse sequence of flat-top pulses. For phase correction, the pulse is separated into three parts, the ramp up/down, flat and gap parts, represented by red, green, and blue colors in the figure. }
    \label{fig:pulse_phase}
\end{figure}

When using a fixed-frequency pulse, each pulse can be separated into two parts, as shown in Figure \ref{fig:pulse_phase}. From $t_i$ to $t_i + t_{\text{gate}}$ is the time the pulse is applied and from $t_i + t_{\text{gate}}$ to $t_{i+1}$ is the time gap between two pulses. During the time gap between two pulses, the detuning between the drive frequency and the qubit frequency is constant. Therefore, $\vph_i$ can be written as
\begin{equation}
    \vph_i = \vph_{\text{gate}_i} + \vph_{gap}, \text{ with } \vph_{gap} = (\omega_{q} - 3 \omega_{d})t_\text{gap}.
    \label{eq:Phase_i}
\end{equation}
For flat-top pulses, $\vph_i$ can be further simplified. By choosing the drive frequency, the accumulated phase of a gate is zero during the flat part and only depends on the edges, which means $\vph_{\text{gate}_i}=\vph_{\text{ramp}_i}$.
When only changing the length of the flat part of the pulse, the $\vph_{\text{gate}_i}$ is a constant for all pulses and only two parameters, $\vph_\text{ramp}$ and $\delta\omega=\omega_{q}-3\omega_{d}$, need to be calibrated.

Frequency-modulated pulses keep the driving frame locked to the qubit frame and all pulses are treated the same way. However, the locking between two frames can drift because of the imperfect pulse shape, error in modulating frequency etc. Therefore, to achieve high-fidelity control, a compensation phase is usually required after each pulse.

Performing a gate in the y-axis direction correctly is also important. After correcting the phase caused by the ac-Stark shift, the remaining phase that needs to be considered is the phase of the gate itself. For example, a $Y$ gate has a phase of $\pi/2$, and a $-X$ gate has a phase of $\pi$. Because the gate is a three-photon transition, only 1/3 of the gate's phase needs to be applied to the drive. For the three-photon subharmonic gates, a trick that can be done is flipping the direction of the y-axis of the drive frame and adding the gate phase directly to the drive.

\subsection{Suppressing leakage to high excitation states}
\label{ssec:DRAG}
Leakage to  non-computational states can caused by activating the leakage directly by crossing through an unwanted transition (i.e. $gf$/6) and due to the usual competition between the drive-induced Rabi rate and the finite anharmonicity of the transmon.
Activating the leakage process directly can be avoided by modulating the pulse frequency based on the Stark shift, as shown in Fig. 4.

Although the subharmonic transitions are closer to each other than the distance between resonant transitions, the leakage can be less serious than resonant driving for the same pulse shape. This is because the Rabi rate of the subharmonic process is proportional to drive amplitude cubed for leakage through the $ef/3$ process and drive amplitude to the sixth power for the $gf/6$ process, and thus is far narrower in frequency spaces/falls off resonance far more quickly.  

To further reduce leakage, DRAG correction can be applied~\cite{motzoi2009simpleSM}. Only considering the first three lowest levels of the transmon qubit, the truncated Hamiltonian of Eq.~\ref{eq:SH_hamiltonian} is
\begin{equation}
    \hat{H}^R/\hbar = (2\alpha|\eta|^2-3\delta)\ket{e}\bra{e} + (\alpha + 4\alpha|\eta|^2-6\delta)\ket{f}\bra{f} + \frac{\alpha}{3}[\eta^{*3} (\ket{g}\bra{e}+\sqrt{2}\ket{e}\bra{f}) + h.c.].
    \label{eq:Hamiltonian_DRAG}
\end{equation}
Defining
\begin{equation}
    \sigma_{j,k}^x =\ket{j}\bra{k} + \ket{j}\bra{k}, \sigma_{j,k}^y = -i\ket{j}\bra{k} + i\ket{k}\bra{j} 
\end{equation}
\begin{equation}
    \zeta^x = Re(\eta^3), \zeta^y = Im(\eta^3),
    \label{eq:zeta}
\end{equation}
the Hamiltonian can be simplified to
\begin{equation}
    \hat{H}/\hbar = (2\alpha|\eta|^2-3\delta)\ket{e}\bra{e} + (\alpha + 4\alpha|\eta|^2-6\delta)\ket{f}\bra{f} + \frac{\alpha}{3}[\zeta^x(\sigma_{g,e}^x + \sqrt{2}\sigma_{e,f}^x) + \zeta^y(\sigma_{g,e}^y + \sqrt{2}\sigma_{e,f}^y)].
\end{equation}

 To implement DRAG, the adiabatic transformation
 \begin{equation}
     V(t) = \text{exp}[-i\frac{\zeta^x}{3}(\sigma_{g,e}^y + \sqrt{2}\sigma_{e,f}^y)]
 \end{equation}
 is performed on the Hamiltonian in Eq.~\ref{eq:Hamiltonian_DRAG}. Assuming $\eta,\ \zeta^x,\ \zeta^y \ll 1$ and $\delta\ll\alpha$, we have
 \begin{equation}
     \begin{split}
         \hat{H}^V \approx&\  
         (2\alpha|\eta|^2 - 3\delta + 
         \frac{2}{9}\alpha\zeta^{x2})\ket{e}\bra{e}+
         (\alpha + 4\alpha|\eta|^2 - 6\delta + \frac{4}{9}\alpha\zeta^{x2})\ket{f}\bra{f}\\
         &+\frac{1}{3}\alpha\zeta^x \sigma_{g,e}^x -
         \frac{\sqrt{2}}{3}(2\alpha|\eta|^2 - 3\delta)\zeta^x\sigma_{e,f}^x+
         \frac{\sqrt{2}}{18}\zeta^{x2}\sigma_{g,f}^x +
         \frac{1}{3}(\alpha\zeta^y + \Dot{\zeta}^x)(\sigma_{g,e}^y+\sqrt{2}\sigma_{e,f}^y).
     \end{split}
 \end{equation}
It shows that to cancel the transition to $\ket{f}$ and ac-Stark shift to the $1^{st}$ order during the drive,
\begin{equation}
    \begin{split}
        &\zeta^y = \frac{-\dot{\zeta}^x}{\alpha},\\
        &\delta = \frac{2}{3}\alpha|\eta|^2-\frac{2}{27}\alpha\zeta^{x2}.
    \end{split}
\end{equation}
Using the definition in Eq.~\ref{eq:zeta}, we can find the correction pulse amplitude by solving the differential equation:
\begin{equation}
    \begin{split}
        &\mathrm{Im}[\eta(t)^3]+\frac{1}{\alpha}\frac{\partial \mathrm{Re}[\eta(t)^3]}{\partial t}=0\\
        \Rightarrow\;\;&\dot{\vep}_x(\vep_x^2-\vep_y^2)-6\vep_x\vep_y\dot{\vep}_y + \alpha(3\vep_x^2\vep_y-\vep_y^3)=0.
    \end{split}
    \label{eq:DRAG_ODE}
\end{equation}

\begin{figure}[ht]
    \centering
    \includegraphics{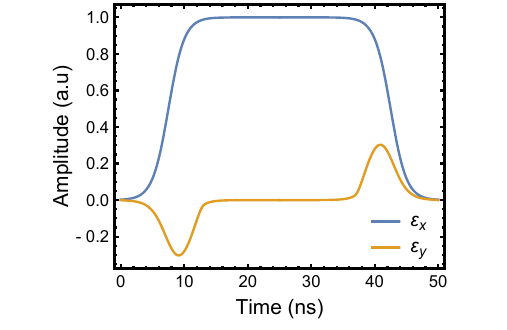}
    \caption{DRAG correction of a \unit[50]{ns} flat-top pulse. The pulse with zero phase and without DRAG correction is defined to only have $\vep_x$ component, shown as the blue trace. The orange trace, $\vep_y$ component, is the DRAG correction of the pulse.}
    \label{fig:DRAG_flattop}
\end{figure}

A pulse with zero phase only has the $\vep_x$ component before applying DRAG correction. Therefore, we can find the correction amplitude by solving for $\vep_y$ in Eq.~\ref{eq:DRAG_ODE}. An example of a \unit[50]{ns} pulse with first-order DRAG correction is given in Figure~\ref{fig:DRAG_flattop}. Pulses with phase $\vph$ can be derived from the \emph{x} and \emph{y} components of zero phase pulse, $\vep_x$ and $\vep_y$,
\begin{equation}
    \begin{split}
        \vep_x' &= \vep_x\text{cos}(\vph/3) - \vep_y\text{sin}(\vph/3)\\
        \vep_y' &= \vep_x\text{sin}(\vph/3) + \vep_y\text{cos}(\vph/3).
    \end{split}
\end{equation}

\subsection{Modeling drive strength}
\label{ssec:drive_strength}
\begin{figure}[ht]
    \centering
    \includegraphics[width=0.9\linewidth]{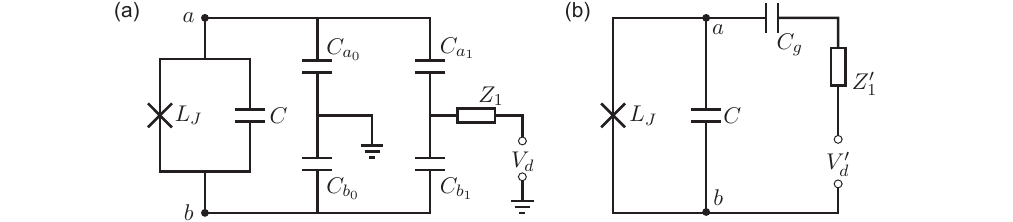}
    \caption{(a) The lumped-element circuit diagram of transmon with single drive port. (b) the reduced equivalent circuit.}
    \label{fig:lumped_circuit}
\end{figure}

As shown in Figure \ref{fig:schematic}, we consider a transmon qubit placed inside an Aluminum tube and only couples to a driving port. It can be modeled by the lumped-element circuit, shown in Figure \ref{fig:lumped_circuit}. The two antenna pads A and B of the transmon are each capacitively coupled to the ground, the wall of the package, and the center pin of the driving port. The Hamiltonian of interaction between transmon and external drive can be written as $\hat{H}_{int} = 2e\hat{n}\beta V_d$, where $\hat{n}$ is the number operator of cooper pairs and $\beta$ is a scaling factor between the voltage at the drive port and the voltage across the transmon's Josephson junction. When the coupling strength between the transmon and the external environment is weak, meaning $C_{a0}, C_{a1}, C_{b0}, C_{b1}\ll1/(Z_1\omega)$, the factor $\beta$ is defined as follow\cite{PechalThesisSM},

\begin{subequations}
    \begin{align}
        C_g &= \frac{(C_{a0} + C_{a1}) (C_{b0} + C_{b1})}{C_{a0} + C_{a1}+C_{b0} + C_{b1}}\\
        Z_1'&=\frac{(C_{a0}+C_{b0})(C_{a1}+C_{b1})}{(C_{a0} + C_{a1})(C_{b0} + C_{b1})}Z_1\\
        Z_\text{tr} &= \frac{Z_{L_J}Z_C}{Z_{L_J}+Z_C}\\
        V_d' &= \frac{C_{a0}C_{b1}-C_{a1}C_{b0}}{(C_{a0}+C_{a1})(C_{b0}+C_{b1})}V_d\\
        \beta &= \frac{V_{ab}}{V_d}=\frac{Z_\text{tr}}{Z_\text{tr}+Z_1'+Z_{Cg}}\frac{V_d'}{V_d}.
    \end{align}
\end{subequations}

Knowing the power sent by room temperature electronics and the total attenuation on the input line, the drive port voltage $V_d$ can be calculated easily.

The Hamiltonian of a qubit under a drive with frequency of $\omega_d$ and amplitude $V$ can be written as,
\begin{equation}
    \begin{split}
        \hat{H} &= \hat{H}_0 + \hat{H}_{int}\\
        &=\frac{1}{2}\hbar\omega_q\sigma_z + 2e\hat{n}\beta V\text{cos}(\omega_dt),
    \end{split}
\end{equation}
where $\hat{n} = n_0\Sigma\sqrt{i+1}(\ket{i}\bra{i+1} + \ket{i+1}\bra{i})$, $n_0=\sqrt[4]{E_j/32E_c}$.\cite{koch2007chargeSM} Since we are only considering the first two energy levels of the transmon, $\hat{n}$ can be simplified to $\hat{n} = n_0\hat{\sigma}_x$. So $\hat{H} = \hbar\omega_q\hat{\sigma}_z/2 + 2en_0\beta V\text{cos}(\omega_dt)\hat{\sigma}_x$

In the interaction picture we have
\begin{equation}
    2en_0\beta V \begin{pmatrix}
        0 & e^{-i\omega_qt}\text{cos}(\omega_dt)\\
        e^{i\omega_qt}\text{cos}(\omega_dt) & 0
    \end{pmatrix}
    \begin{pmatrix}
        c_1(t)\\
        c_0(t)
    \end{pmatrix}=i\hbar
    \begin{pmatrix}
        \dot{c_1}(t)\\
        \dot{c_0}(t)
    \end{pmatrix},
\end{equation}
Assuming $\omega_d = \omega_q$ and taking the rotating wave approximation, we have
\begin{equation}
    en_0\beta V\begin{pmatrix}
        0 & 1\\
        1 & 0
    \end{pmatrix}
    \begin{pmatrix}
        c_1(t)\\
        c_0(t)
    \end{pmatrix}=i\hbar
    \begin{pmatrix}
        \dot{c_1}(t)\\
        \dot{c_0}(t)
    \end{pmatrix},
\end{equation}
It can be solved that
\begin{equation}
    \begin{cases}
        c_1(t) = a_1e^{-igt} + a_2e^{igt}\\
        c_0(t) = b_1e^{-igt} + b_2e^{igt}
    \end{cases},
\end{equation}
where $g=en_0\beta V(t)/\hbar$ and coefficients $a_1$, $a_2$, $b_1$, $b_2$ depend on the initial state $\ket{\vph} = c_1(0)\ket{1}+c_0(0)\ket{0}$. Assuming the initial state is $\ket{\vph} = \ket{0}$, we have $c_1(t) = \text{sin}(gt)e^{i\theta_1}$, $c_0(t) = \text{cos}(gt)e^{i\theta_0}$. The expectation value of $z$-axis measurement is $|c_1|^2-|c_0|^2 = -\text{cos}(2gt)$. So the Rabi rate of the process is

\begin{equation}
    \Omega = 2en_o\beta V/\hbar .
\end{equation}

The Rabi rate of subharmonic process can be calculated together with Eq. \ref{eq:Hamiltonian_RWA},

\begin{equation}
    \Omega_{\text{sub}} = \frac{1}{3}\alpha(\frac{\omega_d \Omega}{\omega_q^{'2}-\omega_d^2})^3.
\end{equation}

\subsection{Subharmonic driving vs. resonant driving}
\label{ssec:sub_res_comparison}
The cooling power of the dilution refrigerator at base temperature ($\sim \unit[20]{mK}$) is limited, usually on the order of $\unit[10]{\mu W}$ due to the low heat conductivity between liquid helium and the metal surface at cryogenic temperature.
Therefore, heat dissipation at the base stage is an important factor to consider when controlling a quantum system.
Meanwhile, the higher-temperature stages have much better cooling capacity, which can handle more driving power before it reaches the limit.
Therefore, here, we compare the heat dissipated at the base stage between resonant drive and subharmonic drive.
In this section, we explain the conditions for comparing the driving methods and the simulation method.

Considering the base stage as an one-port device, the dissipated heat inside the device is determined by the input driving power $P_B$ and the total attenuation on the base stage's drive lines $T_B$ (\unit[10]{dB} attenuation corresponds to $T_B = 0.1$). With the majority of the power being absorbed by the attenuators, the heat dissipated at the base stage is approximately equal to $P_{B}(1-T_{B}^2)$. To compare the performance of the two driving schemes, we want to know that, at same gate speed and $T_1$ limit, which method dissipates more heat. Therefore, we need to find the relation between driving power $P_{B}$ and Rabi rate $\Omega$ under a given coupling strength of the driving port $\kappa_\text{ext}$.

The Rabi rate of the qubit under certain driving power can be calculated using a lumped circuit model, as discussed in Sec.~\ref{ssec:drive_strength}. To extract parameters of the lumped circuit model, we used the FEM software Ansys Maxwell to calculate the capacitance matrix of all conductors, including two antenna pads of the transmon, the tube wall and the probe pin of the drive port. The readout resonator is not implemented in the simulation because it has little effect on qubit driving.

\begin{figure}[ht]
    \centering
    \includegraphics{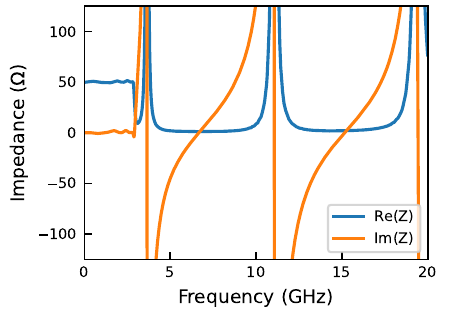}
    \caption{Impedance vs. frequency of low-pass filter used in the experiment}
    \label{fig:LPF_impedance}
\end{figure}

The basic properties of the qubit mode, such as frequency and lifetime $T_1$, are estimated using Ansys HFSS. To model the qubit photon decaying through the port, a lumped RLC port is assigned to the driving port.
For resonant driving, the port is assigned a \unit[50]{$\Omega$} impedance for perfect impedance match. For subharmonic driving, to evaluate the effect of the low-pass filter, the impedance of the port is calculated from the S-parameter measured by VNA at room temperature.
The commercial low-pass filter Mini-Circuits ZLSS-A2R8G-S+ is measured and used in this experiment.
To compare the performance of two driving methods, the coupling port strength is adjusted so that, in both cases, the qubit mode has the Q factor of $2.5\times10^7$, which corresponds to $T_1 \approx \unit[1]{ms}$.

\subsection{Qubit relaxation through multi-photon emission process}
\label{ssec:multi_decay}

\input{supp_decay_v2}

\section{Experimental detail}

\subsection{Qubit characterization}
\label{ssec:qubit_characterization}
Four qubits fabricated on the same wafer were cooled down and characterized in this experiment. Table~\ref{tab:qubit_setup} lists the transmons' coherence times and subharmonic drive line configurations. The five numbers in the attenuation column represent the attenuations on the 77K, 4K, still, 100mK and base stages of the dilution refrigerator. Three models of lowpass filter with cutoff frequency around \unit[2-3]{GHz} were implemented on the drive lines.
Although the subharmonic drive lines ($Q_1$ to $Q_3$) are less attenuated than the regular drive line ($Q_4$), the coherence times, both $T_1$ and $T_2$, are improved by a few times. This shows that the lowpass filters successfully suppress the photon decay through the drive port and reduced the noise seen by the transmon. 

\begin{table}[ht]
    \centering
    \begin{tabular}{c c c c c c c}
         \hline
         & Frequency (MHz) & $T_1$ ($\mu s$) & $T_{2R}$ ($\mu s$) & $T_{2E}$ ($\mu s$) & Attenuation (dB) & Lowpass filter\\
         \hline
         $Q_1$ & 4236.57 & 41 & 21 & 72 & 0, 20, 0,  0, 30 & Mini-Circuits ZLSS-A2R8G-S+\\
         $Q_2$ & 3846.84 & 26 & 33 & 43 & 0, 20, 0,  0, 30 & Mini-Circuits ZLSS252-100W-S+\\
         $Q_3$ & 4449.87 & 42 & 48 & 67 & 0, 20, 0, 20, 20 & Mini-Circuits ZX75LP-2000-S+\\
         $Q_4$ & 4889.54 & 12 &  7 & 23 & 0, 20, 0, 20, 40 & None\\
        \hline
    \end{tabular}
    \caption{Transmon coherence time and the configuration of subharmonic drive lines.}
    \label{tab:qubit_setup}
\end{table}

\begin{figure}[ht]
    \centering
    \includegraphics[width=0.4\linewidth]{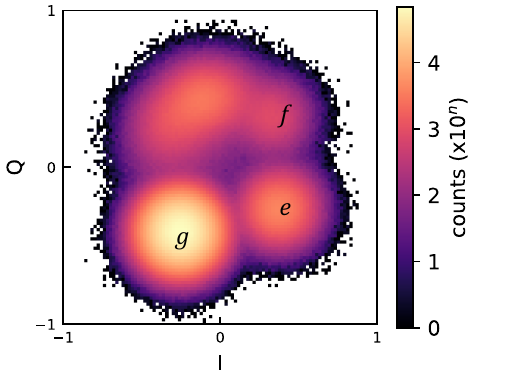}
    \caption{The histogram of transmon readout results.}
    \label{fig:qubit_histogram}
\end{figure}

Quantum limited amplifiers~\cite{kaufman2024simple} are used in this experiment to improve readout fidelity.Flat-top pulses that are \unit[1]{$\mu$s} long are used to read out the qubit. The histogram shows that we can discriminate states $\ket{g}$, $\ket{e}$, $\ket{f}$, as shown in Fig.~\ref{fig:qubit_histogram}.
\subsection{Amplitude Calibration}
\label{ssec:amp_calibration}
\begin{figure}[ht]
    \centering
    \includegraphics[width=0.5\linewidth]{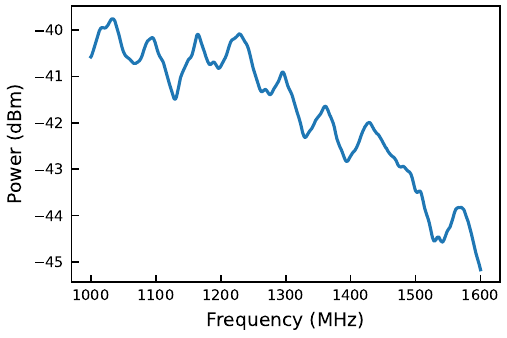}
    \caption{The microwave power at base stage vs. drive frequency.}
    \label{fig:line_calibration}
\end{figure}

The controlling pulses are generated by direct digital synthesis (DDS) with the Xilinx RFSoC ZCU216. Due to the property of DACs, filters, and other microwave components used on the subharmonic drive line, the insertion loss of the drive line is not constant but frequency dependent, as shown in Fig.~\ref{fig:line_calibration}.  The power in Fig.~\ref{fig:line_calibration} is measured with a spectrum analyzer connected to the base stage of the dilution refrigerator and at a fixed digitally defined DAC amplitude.

\begin{figure}[ht]
    \centering
    \includegraphics{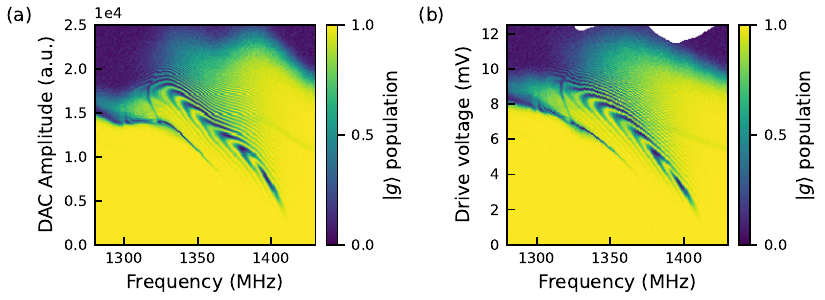}
    \caption{The power Rabi experiment (a) without amplitude correction and (b) with amplitude correction.}
    \label{fig:rabi_correction}
\end{figure}

Figure~\ref{fig:rabi_correction} shows the power Rabi experiment with and without amplitude correction. By applying the correction, the distortion of the Rabi fringes is mostly removed. The amplitude calibration is also important to find the chirp modulation frequency.

\subsection{Pulse tune-up}
\label{ssec:pulse_tuneup}
To perform single-qubit control, we need to be able to perform $\pi$ pulses and $\pi/2$ pulses in the $X$ and $Y$ directions. For each gate, as shown in Eq.~\ref{eq:flat-top},
a flat-top waveform is defined with four free parameters : amplitude $A_0$, pulse length $t_0$, ramp speed $k$ and cutoff factor $t_1/k$. In addition to pulse shape, frequency modulation can be applied through $\varphi(t)$.

\begin{figure}[ht]
    \centering
    \includegraphics[width=1\linewidth]{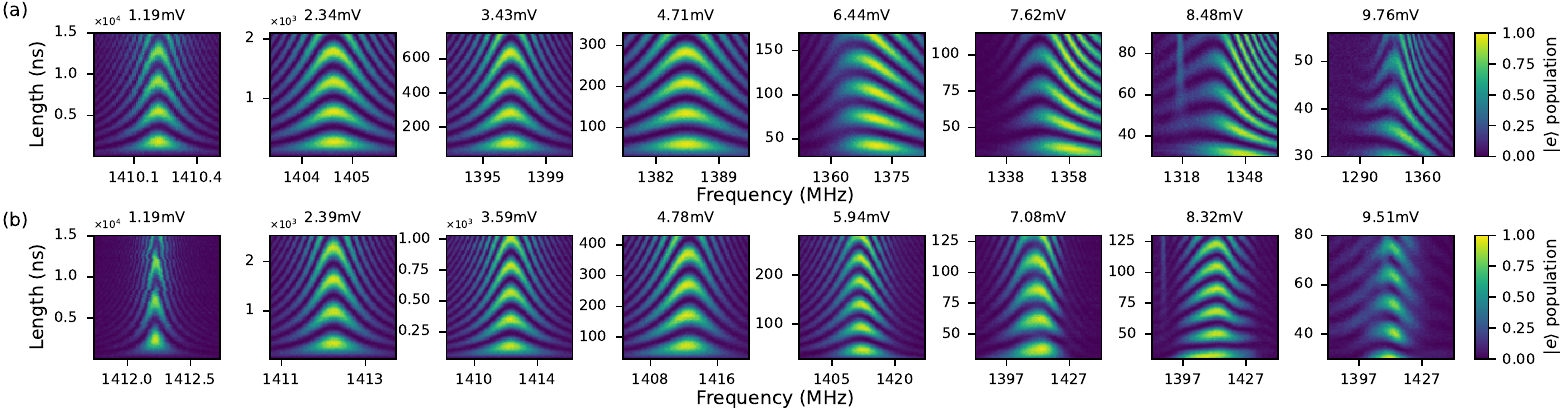}
    \caption{Subharmonic Rabi oscillations under different driving amplitudes with
    (a) the fixed frequency pulse and 
    (b) the chirp modulated pulse.}
    \label{fig:time_rabi_all}
\end{figure}

The pulse tune-up procedure starts from the Rabi experiment, which finds the desired drive strength and the corresponding drive frequency. Fig.~\ref{fig:Rabi} in the main paper is a summary of multiple Rabi experiments performed with different driving amplitudes and fixed-frequency pulses. The full experimental results are shown in Figure~\ref{fig:time_rabi_all}(a). Each plot is fitted to extract the Rabi rate and AC-Stark shift and corresponds to a pair of data points in Figure~\ref{fig:Rabi}(b). When driven strongly, the Rabi oscillation pattern shows asymmetric distortion due to the large frequency shift during a pulse. To reduce the distortion and avoid leakage to higher excited states, we applied chirp frequency modulation that tracks the transmon's frequency shift during the subharmonic drives. Fig.~\ref{fig:power_rabi_chirp}(b) and Fig.~\ref{fig:time_rabi_all}(b) show the experiment result using frequency-modulated pulses. The frequency-modulated pulses cancel most of the frequency detune during the drive and, as a result, the Rabi pattern distortion is reduced.

From the Rabi experiment, we can pick a drive amplitude $A_0$ for the rest of the pulse tune up procedure. For the pulse envelope, the parameters $k$ and $t_1$ are usually fixed as well, leaving the pulse length the only free parameter to tune up. The length of the $\pi$ pulse and $\pi/2$ pulse is acquired by fitting time-Rabi experiment. For the frequency of the drive, the LO frequency is acquired by measuring the static $g-e$ frequency of the transmon and the modulating frequency is acquired by measuring the transmon's AC-Stark shift. The chirp frequency modulation is defined as
\begin{equation}
    \varphi_\text{chirp}(t) = \int_0^t\omega_\text{chirp}(t')dt' \text{, with }\omega_\text{chirp}(t)=\omega_\text{mod}(A_0) (\frac{A(t)}{A_0})^2.
\end{equation}
The modulation frequency $\omega_\text{mod}$ at amplitude $A_0$ is measured by qubit spectroscopy. However, because the quadratic model does not fully cancel the phase drift, an additional phase needs to be calibrated. It can be measured by playing two $\pi/2$ pulses with a minimum time gap between them. The phase of the second pulse is swept to find the correct compensating phase so that the pulses rotate over the same axis, which puts the final state of the transmon in $\ket{e}$. 

Next, gate sequences that repeat the target gate multiple times are performed to amplify the gate error and fine-tune the pulse. 
\begin{figure}[ht]
    \centering
    \includegraphics{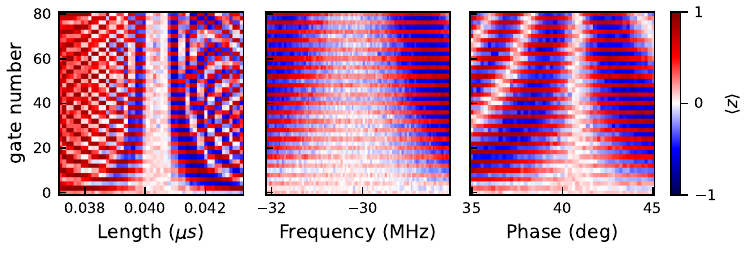}
    \caption{\textbf{Fine tuning gate parameters.} The transmon is prepare into the equator of the Bloch sphere with a $\sqrt{X}$ gate, then followed by $2n$ $\sqrt{X}$ gate, sweeping pulse parameters pulse length $t_0$, modulating frequency $\omega_\text{mod}$ and correcting phase. If the $\sqrt{X}$ is well defined, $\langle z \rangle$ should remain zero.} 
    \label{fig:tune-up}
\end{figure}
These tune-up procedures may be applied iteratively, each time scanning for finer control of the pulse parameters. More calibration steps can be applied if DRAG is required. However, with the gate time currently achieved and the present qubit coherence time, leakage to higher excited states is not the limiting factor of gate fidelity, and so we omit DRAG. Finally, interleaved randomized benchmarking is used to quantitatively measure the fidelity of each gate.

\subsection{Temperature Stabilization}
\label{ssec:temperature}
\begin{figure}[ht]
    \centering
    \includegraphics[width=1\linewidth]{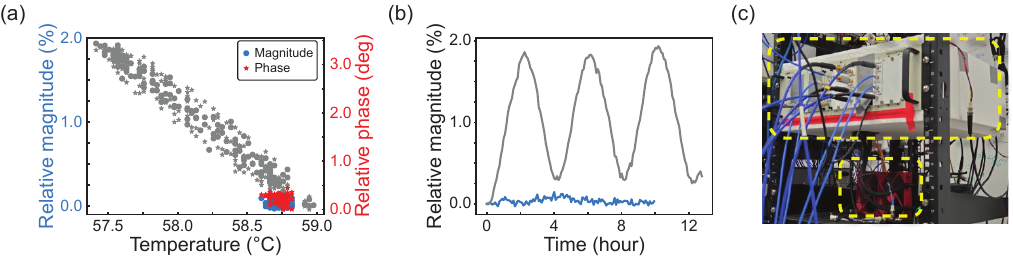}
    \caption{Temperature stability and its effect on pulse amplitude/phase.
    (a) The amplitude vs. temperature before (grey circles and stars) and after stabilization (blue circles and red stars). (b) The amplitude stability over time without stabilization (grey trace) and with stabilization (blue trace). The great improvement in pulse amplitude stability makes high fidelity control possible. (c) The experiment setup and a PID-controlled commercial liquid cooler are highlighted with yellow dashed boxes. The box is wrapped with Styrofoam except the front and back faces, which are for connecting coaxial cables. The liquid cooling head is mounted on the Al plate at the bottom of the box.}
    \label{fig:temp_data}
\end{figure}

See the Methods for the majority of these details.  Data on temperature stability vs. time and an image of the temperature stabilization rig are shown in Fig.~\ref{fig:temp_data}.

\bibliography{refs_SM}

%% file: supp_decay_v2.tex
\newcommand{\vac}{\vert \text{vac} \rangle \langle \text{vac} \vert}

In this sub-section, we calculate the decay rate of a transmon qubit via sub-harmonic relaxation. We specifically focus on the relaxation from the first excited state $\ket{1}$ to the ground state $\ket{0}$ via the emission of three photons, each carrying away approximately $\hbar \omega_q/3$ of energy.

The Hamiltonian that describes the system (i.e. the transmon qubit) and the photon bath (i.e. the transmission line) is,
\begin{align}
    H \quad & = H_{\text{sys}} + H_{\text{bath}} + H_{\text{s-b}} \nonumber \\
    H_{\text{sys}} & = \omega_q q^{\dagger} q + H_{\text{nl}} \\
    H_{\text{bath}} & = \int d\nu \, \nu \, b_{\nu}^{\dagger} b_{\nu}  \\
    H_{\text{s-b}} & = \int d\nu \lambda(\nu) (b_\nu^\dagger q + b_\nu q^\dagger )
\end{align}
where we set $\hbar = 1$, $b^\dagger_\nu$ creates a bath photon with angular frequency $\nu$ and $\lambda(\nu)$ describes the system-bath coupling and we define its form below. Terms contributing to $H_\text{nl}$, which describe the nonlinearity of the transmon, can be found in Eq.~\eqref{eq:transmon_h}. We remark that the terms $q^4$ and $(q^\dagger)^4$ do not participate in the lowest order dynamics, and hence we ignore these two terms for simplicity. We specifically focus on the nonlinear Hamiltonian 
\begin{align}
    H_{\text{nl}} =g_2 (q^\dagger q^\dagger + q q) +  g_4 (q^\dagger q^\dagger q^\dagger q + q^\dagger q q q) + g_4' q^\dagger q^\dagger q q,
\end{align}
where we define $g_2 = \alpha/2$, $g_4 = \alpha/3$ and $g_4' = \alpha /2$, and $\alpha$ is the anharmonicity of the qubit. The system-bath coupling strength is
\begin{equation}
    \lambda(\nu) = \Theta(\nu) \frac{C_{c}}{\sqrt{C_r c}} \sqrt{\frac{\omega_0 \nu}{2 \pi v_{\text{tl}}} },
    \label{eq:coupling}
\end{equation}
where $C_{c}$ is the coupling capacitance between the qubit and the transmission line, $C_{r}$ is the capacitance of the qubit, $c$ is the characteristic capacitance of the transmission line, $v_{\text{tl}}$ is the speed of the transmission line, $\omega_0$ is the frequency of the qubit~\cite{blais2021circuitSM}. In our case, we put a cut-off (filter) function $\Theta(\nu)$ to suppress the high-frequency system-bath coupling. Specifically, with the Kerr nonlinearity, we assume
\begin{align}
    \Theta(\nu) = \left\lbrace
    \begin{array}{cl}
        1 & \vert \nu \vert \leq \omega_0/3 + \vartheta \\
        0 & \vert \nu \vert > \omega_0/3 + \vartheta
    \end{array}
    \right.
\end{align}

Due to the energy conservation, we consider a three-photon relaxation process, i.e., a single transmon excitation decays to three bath photons. The lowest-order term contributing to the decay rate of the qubit comes from fourth-order Fermi's golden rule perturbation theory~\cite{sakurai2011modern}. Specifically we are looking for transitions from the initial state $\ket{1; \text{vac}}$ to the final state $\ket{0; 1_{\nu_1}, 1_{\nu_2},1_{\nu_3}}$, where the first half of the ket notation indicates the state of the qubit and the second part the state of the bath. We ignore the transitions that end up with fewer than three photons, as these transitions are suppressed by the system-bath low-pass filter. We remark that transitions that end up with three photons in fewer than three modes are possible and do contribute to the decay rate. However, the rate for these transitions scales as $(D(\nu)^{-1}/\omega_0)$ (to two distinct modes) and $(D(\nu)^{-1}/\omega_0)^2$ (all photons go into the same mode), where $D(\nu)$ is the density of states in the transmission line. The density of states of a length $l$ transmission line with dispersion relation $\nu = v_{\text{tl}} k$ is given by
\begin{equation}
    D(\nu) = \frac{l}{2\pi \nu_{\text{tl}}}.
    \label{eq:dof}
\end{equation}
Specifically, a $1$~m long transmission line with $v_{\text{tl}} \sim c/1.5$, the density of states $D(\nu) \sim 0.032~\text{GHz}^{-1}$, which gives $D(\nu)^{-1}/\omega_0 \sim 0.01$. This ratio is small compared to other frequency ratios (e.g. $\vert \alpha \vert / \omega_0 \sim 0.1$ order). Therefore, we ignore these transitions. Here we also assume the system-bath coupling bandwidth $\vartheta/\omega_0$ is around the order $0.1$, which is of the same order of $\vert \alpha \vert / \omega_0$.

The transition rate can then be calculated from Fermi's golden rule as 
\begin{equation}
    \begin{aligned}
    \Gamma = 2 \pi \frac{1}{6} \int_{\omega_0/3-\vartheta}^{\omega_0/3+\vartheta} & d \nu_1 d \nu_2 d \nu_3 \delta(\omega_0 - \nu_1 - \nu_2 - \nu_3) \\
    & \times \left\vert 
        \sum_{m,n,p} 
          \frac{\bra{0; 1_{\nu_1}, 1_{\nu_2},1_{\nu_3}} V \ketbra{m} V \ketbra{n} V \ketbra{p} V \ket{1; \text{vac}}}
          {(\omega_0 - \varepsilon_{p} + i \eta) (\omega_0 - \varepsilon_{n} + i \eta) (\omega_0 - \varepsilon_{m} + i \eta)}
    \right\vert^2,
\end{aligned}
\end{equation}

where $V = H_{\text{nl}} + H_{\text{s-b}}$, the summation is over all possible system-bath states, $\varepsilon_{m,n,p}$ are the energies of the system-bath state $\ket{m}$, $\ket{n}$ and $\ket{p}$, the factor of $1/6$ is to remove the over-counting induced by the frequency integrals, and we take the limit $\eta \rightarrow 0_+$. We then count the virtual paths that are involved in the fourth-order perturbation. Note that there are two distinct sets of possible paths. One set of the paths consists of states 
\begin{align}
    \ket{p} = \ket{3;\text{vac}}, \quad 
    \ket{n} = \ket{2; 1_{\nu_1}},  \quad 
    \ket{m} = \ket{1; 1_{\nu_1}, 1_{\nu_2}},  \quad 
    \label{eq:path1}
\end{align}
and all possible permutations of $\nu_1$, $\nu_2$, and $\nu_3$. The other set of paths includes the path
\begin{align}
    \ket{p} = \ket{0; 1_{\nu_1}}, \quad 
    \ket{n} = \ket{2; 1_{\nu_1}},  \quad 
    \ket{m} = \ket{1; 1_{\nu_1}, 1_{\nu_2}},  \quad 
    \label{eq:path2}
\end{align}
and all the other paths formed by permuting $\nu_1$, $\nu_2$, and $\nu_3$. Putting these considerations together, we obtain the three-photon relaxation rate 
\begin{align}
    \Gamma & = \frac{2\pi}{6} \int d\nu_1 d \nu_2 d \nu_3 \delta(\omega_0 - \nu_1 - \nu_2 - \nu_3)  \nonumber \\
    & \times \Bigg\vert 
        \sum_{\{\nu_1, \nu_2, \nu_3\}} \Bigg( 
        \frac{6\lambda(\nu_1) \lambda(\nu_2) \lambda(\nu_3) (g_2 + g_4)}{(-\nu_1 - \nu_2 +i \eta)(-\omega_0 - \nu_1 + i \eta)(-2\omega_0 +i \eta)} \nonumber \\
    & \qquad + \frac{2\lambda(\nu_1) \lambda(\nu_2) \lambda(\nu_3) g_2}{(-\nu_1 - \nu_2 +i \eta)(-\omega_0 - \nu_1 + i \eta)(\omega_0 -\nu_3 +i \eta)} 
    \Bigg)
    \Bigg\vert^2,
\end{align}
where $\sum_{\{\nu_1, \nu_2, \nu_3\}}$ indicates a summation over all terms that are generated by permuting $\nu_1$, $\nu_2$, and $\nu_3$. We then make the further assumption that $\vartheta/\omega_0 \sim \vert \alpha \vert / \omega_0 \ll 1$. We find that the relaxation rate, to lowest order in $\vartheta/\omega_0$, is
\begin{align}
\label{eq:normal_decay_rate}
    \Gamma = \frac{243}{32 \pi^2} \frac{\gamma_1^3 \vert \alpha \vert^2}{\omega_0^4} \left( \frac{\vartheta}{\omega_0}\right)^2,
\end{align}
where we consider that in the coupling bandwidth, the system-bath coupling is slowly varying in the band with $\lambda(\nu) \sim \lambda(\omega_0/3)$, and we define $\gamma_1 = 2\pi \vert \lambda(\omega_0/3) \vert^2$. 

We further note that when the transmon is coupled to a thermal bath, which has a mean thermal excitation number 
\begin{align}
 \Tr_{\text{bath}} \left( b_{\nu_1} b_{\nu_1}^\dagger \rho_{\text{b}} \right) = (\bar{n}_{\nu_1}+1),
\end{align}
the relaxation rate will depend on the thermal bath temperature. Because the three-photon relaxation process requires coupling the transmon to three bath photons, the thermal relaxation rate is
\begin{align}
    \Gamma_{\text{th, relax}} = (\bar{n}_{\omega_0/3} + 1)^3 \Gamma. 
\end{align}
This can be seen from the fact that when calculating the jump term in the master equation, we will have terms in the form of $\Tr{b_{\nu_1}^\dagger b_{\nu_2}^\dagger b_{\nu_3}^\dagger \rho_{\text{b}}b_{\nu_3} b_{\nu_2} b_{\nu_1}}$. Meanwhile, the bath will thermally excite the transmon with a rate
\begin{align}
    \Gamma_{\text{th, excite}} = \bar{n}_{\omega_0/3}^3 \Gamma.
\end{align}

 In the case of the transmon coupled to the subharmonic driving field, in addition to the normal relaxation path, the drive also induces addition relaxation paths. Let's consider the terms
\begin{equation}
    H_{s, \text{ap}} = \omega a^\dagger a + g_4' \eta e^{- i \omega_0 t} a^\dagger a^\dagger a + h.c.
\end{equation}
Going into the frame rotating at $\omega_0$, the Hamiltonian can be transformed as
\begin{equation}
    H_{s, \text{ap}} = (\omega-\omega_0) a^\dagger a + g_4' \alpha_{-}^*  a^\dagger a^\dagger a + h.c.
\end{equation}
When we consider the relaxation from the $\ket{e}$ state to the $\ket{g}$ state, this is equivalent to having a qubit excitation with frequency $(\omega - \omega_0) = 2 \omega_0$, and relax to bath photons centered around the $\omega_0$ frequency. Therefore, this can give us a lower-order transition as $\ket{e} \ket{\text{vac}} \rightarrow \ket{g} \ket{1_{\omega}, 1_{\omega'}}$. The effective Hamiltonian matrix element in this case is
\begin{equation}
    V_{i \rightarrow f}^{(2)} \sim \frac{g_4 '\lambda(\nu)^2 \eta}{\omega^2} \sim \frac{g_4' \lambda(\nu)^3}{\omega^3} \cdot \frac{\omega \eta}{\lambda(\nu)} \sim \frac{g_4' \lambda(\nu)^2}{ \omega^3} (\lambda(\nu) \varepsilon).
\end{equation}
We can then estimate the decay rate as
\begin{align}
    \Gamma^{(2)} & \sim \int_{\vartheta} d \omega_{1,2} \vert V_{i \rightarrow f}^{(2)} \vert^2 D^2(\omega) \delta(\omega_i - \omega_f)  \\
    &  \sim \frac{g_4'^2 \gamma_1^3 \vartheta}{\omega^6} \frac{P_d}{\hbar \omega}
\end{align}
Comparing to the normal loss rate Eq.~\eqref{eq:normal_decay_rate}, the decay rate here is
\begin{equation}
    \Gamma^{(2)} \sim \Gamma \frac{P_d}{\hbar \omega \vartheta}.
\end{equation}